\documentclass[12pt,a4paper]{article}

\usepackage{amssymb,amsmath,amsfonts,nicefrac}
\usepackage{graphicx}
\usepackage{slashed}
\usepackage{textcomp}
\usepackage[numbers,sort&compress]{natbib}
\usepackage{psfrag}
\usepackage[dvips,bookmarks,bookmarksnumbered=true,colorlinks=true,linktocpage=true]{hyperref}

\newcommand{\mpi}{M_{\pi}}
\newcommand{\fpi}{F_{\pi}}
\newcommand{\Mpi}{M_{\pi}}
\newcommand{\Fpi}{F_{\pi}}

\newcommand{\epp}{\varepsilon_\pi}

\newcommand{\lp}{\lambda_\pi}

\newcommand{\be}{\begin{equation}}
\newcommand{\ee}{\end{equation}}
\newcommand{\beq}{\begin{equation}}
\newcommand{\eeq}{\end{equation}}
\newcommand{\bea}{\begin{eqnarray}}
\newcommand{\eea}{\end{eqnarray}}
\newcommand{\bdm}{\begin{displaymath}}
\newcommand{\edm}{\end{displaymath}}

\newcommand{\lbar}[1]{\bar{\ell}_{#1}}

\newcommand{\mr}{\mathrm}

\newcommand{\MeV}{\,\mr{MeV}}
\newcommand{\GeV}{\,\mr{GeV}}

\newcommand{\fm}{\,\mr{fm}}

\newcommand{\nn}{\nonumber\\}
\newcommand{\inv}[1]{\frac{1}{#1}}

\newcommand{\eppb}{\bar \varepsilon_\pi}
\newcommand{\eppt}{\tilde \varepsilon_\pi}
\newcommand{\HM}{\ensuremath{H}}
\newcommand{\HMs}{\ensuremath{H^*}}
\newcommand{\mH}{\ensuremath{m_H}}
\newcommand{\mHs}{\ensuremath{m_{H^*}}}
\newcommand{\gHM}{\ensuremath{g_{\pi \HM \HMs}}}

\def\fs{\; \; .}
\def\co{\; \; ,}

\begin{document}

\title{
\vspace{-2cm}
\begin{flushright}
\small
LU TP 10-12
\end{flushright}
\vspace{2cm}
\LARGE\bf Finite volume effects for nucleon and heavy meson masses}

\author{Gilberto Colangelo$^a$, Andreas Fuhrer$^b$ and Stefan
Lanz$^{a,c}$\\[2mm]
{\small $^a\,$Albert Einstein Center for Fundamental Physics,}\\
{\small Institute for Theoretical Physics, University of Bern,}\\   
{\small Sidlerstrasse 5, CH-3012 Bern, Switzerland }\\
{\small $^b\,$Department of Physics, University of California at San
Diego,}\\
{\small 9500 Gilman Drive, La Jolla, California 92093, USA}\\
{\small $^c\,$Department of Astronomy and Theoretical Physics, Lund University}\\
{\small S\"olvegatan 14A, S 223 62 Lund, Sweden}
}

\maketitle

\begin{abstract}
  We apply the resummed version of the L\"uscher formula to analyze finite
  volume corrections to the mass of the nucleon and of heavy mesons. We
  show that by applying the subthreshold expansion of the scattering
  amplitudes one can express the finite volume corrections in terms of only
  a few physical observables and the size of the box.  In the case of the
  nucleon, the available information about the quark mass dependence of
  these physical quantities is discussed and used to assess the finite
  volume corrections to the nucleon mass as a function of the quark mass
  including a detailed analysis of the remaining uncertainties.  For heavy
  mesons, the L\"uscher formula is derived both fully relativistically and
  in a nonrelativistic approximation and a first attempt at a numerical
  analysis is made.

\end{abstract}

\thispagestyle{empty}
\setcounter{page}{0}
\clearpage

\tableofcontents

\newpage
\section{Introduction}

Lattice calculations have recently almost reached physical pion masses (see
e.g. Refs.~\cite{BMW2,Aoki:2008sm}), thereby establishing firm contact with
chiral perturbation theory (ChPT) \cite{Weinberg,GL1983,GL1985}. The
effective theory therefore becomes increasingly useful to address
systematic effects in lattice calculations like the finite volume, for
example. To use the effective chiral Lagrangian to evaluate such effects
was first advocated in Refs.~\cite{Gasser:1986vb,Gasser:1987zq} where it
was also shown that in the $p$-regime, where the Compton wavelength of the
pion is much smaller than the size of the box, the infinite volume
Lagrangian can be used for doing calculations. An alternative approach to
evaluate finite volume corrections (FVC) to masses in the $p$-regime is the
L\"uscher formula \cite{Luscher:1983rk,Luscher:1985dn}. This relates the FVC
of a particle $P$ to an integral over the forward scattering amplitude of
the same particle $P$ off the lightest particle in the spectrum. The main
contributions to this integral come from the low-energy region, while
high-energy contributions are exponentially suppressed. This justifies the
use of amplitudes evaluated in ChPT as an input. An improved version of the
L\"uscher formula, where subleading terms are resummed, was already
successfully applied to meson masses and, in a generalized form, to
coupling constants
\cite{Colangelo:2003hf,Colangelo:2004xr,Colangelo:2005gd,Colangelo:2010cu}.
FVC for the nucleon mass in ChPT and the L\"uscher formula have been
discussed in \cite{AliKhan:2003cu,Koma:2004wz,CFH}.

One of the advantages of the L\"uscher formula is that it allows a
relatively easy estimate of higher order effects and yields simple
analytical expressions. In Ref.~\cite{Colangelo:2005gd} it was shown that
inserting the tree-level $\pi \pi$ scattering amplitude into the resummed
L\"uscher formula exactly reproduces the one-loop result of the FVC for the
pion mass evaluated in ChPT. At the two-loop level, this does not hold
anymore, because in the derivation of the L\"uscher formula, only one
propagator is considered to be in finite volume --- two-loop contributions
in which propagators appearing in different loops are simultaneously taken
in finite volume are exponentially subdominant and neglected from the start
in L\"uscher's approach. However, the two-loop FVC of the pion mass were
explicitly calculated and compared with the result of the resummed
L\"uscher formula with a one-loop amplitude in the integrand in
Ref.~\cite{Colangelo:2006mp}. It was found that the difference is very
small and thereby a solid argument was provided to base further analyses of
FVC on the resummed L\"uscher formula.

\begin{figure}[t]
\begin{center}
\psfrag{A}[][][1.9]{$\nu$}
\psfrag{B}[][][1.3]{$\Mpi$}
\psfrag{C}{\sf integration path}
\psfrag{D}[l]{$\displaystyle -\frac{\Mpi^2+m_H^2-m_{H^*}^2}{2 m_H}$}
\psfrag{E}[r]{$\displaystyle -\frac{\Mpi^2}{2 m_H}$}
\includegraphics[width=10.5cm]{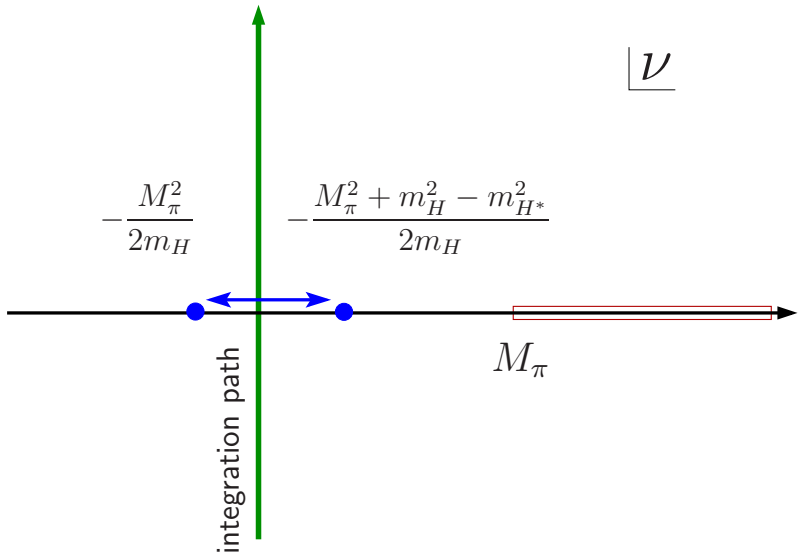}
\end{center}
\caption{\label{fig:nu} Singularities of the forward scattering amplitude
  in the complex $\nu$-plane. 
  The cut on the real axis is due to the intermediate state $H \pi$ and the
  pole on the left-hand side of the imaginary axis to the intermediate
  $H^{*}$ state (or $H$ in the case of the nucleon). The position
  of the pole depends on $m_{H^*}^2 - m_H^2$ and moves to
  the right-hand side of the imaginary axis if $m_{H^*}^2 - m_H^2>\Mpi^2$. 
  Notice that the amplitude is an even function of $\nu$ and that for
  clarity we have drawn only half of the singularities.}
\end{figure}
In this paper we apply the latter to estimate the FVC to masses of heavy
particles, and, in particular, to nucleons and to mesons containing a heavy
quark. Although these two kinds of particles are very different --- nucleons
are spin-$\nicefrac{1}{2}$ particles and are much lighter than the spin-$0$
$B$ mesons, for example --- the essential common feature is that both are
much heavier than the pions. Neither the exact value of their mass, nor
their spin, make a qualitative difference in the evaluation of the FVC.
What makes the treatment of the two cases very similar is that the
singularity structure of the $\pi H$ scattering amplitude at low energy
($H$ being the heavy particle) is dominated by a single pole due to the
exchange of a particle degenerate (or almost degenerate) with $H$. The
forward scattering amplitude is a function of only one kinematical variable
$\nu$. The pole due to the exchange of a particle $H^*$ in the $s$-channel
is located at 
\be
\nu= -(M_\pi^2-m_{H^*}^2+m_H^2)/(2m_H) \stackrel{m_{H^*} \to
  m_H}{\longrightarrow} - M_\pi^2/(2 m_H)\;,
\ee 
as illustrated in
Fig.~\ref{fig:nu}. If $m_{H^*}^2-m_H^2>M_\pi^2$ the pole is on the
right-hand side of the integration path (which coincides with the imaginary
axis) and the L\"uscher formula contains only the term with the integral. As we
take the limit $m_{H^*} \to m_H$ the pole moves to the left until it
crosses the imaginary axis and finally reaches $\nu=- M_\pi^2/(2 m_H)$.
When the pole crosses the imaginary axis an extra local contribution to the
L\"uscher formula appears \cite{Luscher:1983rk}, which is exponentially
leading. The study of the heavy meson case, in which we can vary $\Delta_*
\equiv m_{H^*}-m_H$ and even send it to zero,
allows us to better understand the role of the local,
exponentially leading contribution in the L\"uscher formula.  We stress
that this singularity structure is different from the one in the case of
the light pseudoscalar mesons $L$ considered in
Ref.~\cite{Colangelo:2005gd}. In that case the most important low-energy
singularity is the cut due to the $\pi L$ intermediate state (also shown in
Fig.~\ref{fig:nu} starting at $\nu=M_\pi$ for all particles): the closest
single pole is due to the exchange of a vector meson, but is at higher
$\nu$ values.

For the case of the nucleon we can rely on existing calculations of the
infinite volume $\pi N$ scattering amplitude to one loop and on the
knowledge of the low-energy constants (LEC) appearing therein. In
particular we show how the application of the so-called subthreshold
expansion --- very well known in the phenomenology --- leads to a particularly
simple formula for FVC of the nucleon mass. The solid phenomenological
knowledge of the subthreshold coefficients allows a very precise
determination of the FVC at or near the physical value of the quark mass.
In our numerical analysis we show how the error in the evaluation of the
FVC increases as we go to higher quark masses and identify the LEC which
control these effects. In particular a better understanding of the quark
mass dependence of the axial-vector coupling constant of the nucleon would
lead to a much better control of the FVC. The relevance of this issue for
the calculation of the FVC for the nucleon mass has been already discussed
at length in Ref.~\cite{CFH}, where we presented preliminary results
(dating already five years back). Indeed until lattice calculations of
$g_A$ at lower quark masses were published, we were not able to 
present a reliable numerical analysis of these FVC. The situation has now
clearly improved, but is not yet fully satisfactory and we do think that
there is still a lot to be understood about the behavior of $g_A$ and
other nucleon-related quantities as functions of the quark mass.

Only very few publications on finite volume effects for heavy mesons exist so far
\cite{Goity:1990jb,Arndt:2004bg,Lin:2004xd,Bernardoni:2009sx,Bernardoni:2009um},
most of which are not concerned with the FVC to
the mass. Also about the scattering amplitude of pions off heavy mesons a lot
less is known, even only in the low-energy region. We give
the expression of the tree-level scattering amplitude and show how the
L\"uscher formula needs to be modified if one considers the heavy mesons
away from the infinite mass limit ({\em i.e.} as nondegenerate). We show
that also in this case one can define a subthreshold expansion completely
analogous to the case of the nucleon, and express the FVC in terms of the
subthreshold parameters, so emphasizing the close analogy between the two
cases. We provide a simple numerical analysis on the basis of the limited
information which is available on the $\pi H H^*$ (for $H=D$ or $B$ mesons)
coupling constant and discuss the difference between the fully relativistic
case and the nonrelativistic limit of the L\"uscher formula.

\section{Resummed L\"uscher formula for the nucleon mass}
In this section we present and discuss the formula we will use for
analyzing the finite volume effects on the nucleon mass numerically. We
rely on the resummed version of the L\"uscher formula
\cite{Luscher:1983rk,Luscher:1985dn} which has been proposed in
Ref.~\cite{Colangelo:2005gd}. The formula for the relative finite volume
correction $R_N\equiv(m_N(L)-m_N)/m_N$ reads as follows: 
\bea
R_N\!\!&\!\!=\!\!&\!\! \frac{3 \epp^2 }{4 \pi^2 } \sum_{n=1}^{\infty}
\frac{m(n)}{\sqrt{n} \lp}\left[ 2 \pi \epp g_{\pi N}^2
e^{-\sqrt{n(1-\epp^2)}\lp}- \!  \int_{-\infty}^{\infty} \! dy
e^{-\sqrt{n(1+y^2)}\lp} \tilde D^{+}(y) \right] \; \; \; \; \;
\label{eq:RN}
\eea
where 
\be
\epp=\frac{M_\pi}{2 m_N}\; , \; \; \lp= M_\pi L \; , \; \; \tilde D^{+}(y)=
m_N D^+(i M_\pi y,0) 
\label{eq:defs}
\ee
where $D^+$ is one of the components of the elastic $\pi N$ scattering
amplitude, which is defined as
\be
T(\pi^a(q) N(p) \to \pi^{a'}(q') N(p'))\equiv
T_{a'a}=\delta_{a'a}T^++\frac{1}{2} [\tau_{a'},\tau_a] T^- \fs
\ee
Each of the two isospin components is then broken down into
\be
T^\pm=\bar u'\left[ D^\pm(\nu,t)-\frac{1}{4 m_N} [ q \hskip -0.2cm/',q
  \hskip -0.2cm / ] B^\pm(\nu,t) \right] u
\ee
and each of the amplitudes depends on the kinematical variables $t$ and
$\nu$, defined as 
\be
 t=(q-q')^2 \; ,\; \; \nu=\frac{s-u}{4 m_N} \; , \quad \mbox{where} \quad 
s=(p+q)^2\; , \; \; \; \; u=(p-q')^2 \; .
\label{eq:nutN}
\ee
\begin{table}[b]
\centering
\begin{tabular}{l|rrrrrrrrrrrrrrr}
$n$&1&2&3&4&5&6&7&8&9&10&11&12&13&14&15\\
\hline 
$m(n)$&6&12&8&6&24&24&0&12&30&24&24&8&24&48&0
\end{tabular}
\caption{The multiplicities $m(n)$ for $n\leq 15$.}
\label{tab:m}
\end{table}
The term with $n=1$ in Eq.~(\ref{eq:RN}) has been first given by L\"uscher
\cite{Luscher:1983rk} as an application of his general formula for finite
size corrections to masses \cite{Luscher:1985dn}, but with a factor $2$
missing in the first term within the brackets. This has been pointed out
in Ref.~\cite{AliKhan:2003cu} and later confirmed in
Ref.~\cite{Koma:2004wz} --- it is 
indeed easy to follow the steps in L\"uscher's general proof
\cite{Luscher:1985dn} adapting it to the present case and derive 
formula~(\ref{eq:RN}) as it stands.

The extension to the exponentially suppressed terms with $n>1$ can be
immediately obtained following the general derivation given in
Ref.~\cite{Colangelo:2005gd}. They appear weighted by the multiplicities
$m(n)$ --- these are easily calculable and are reproduced here for
convenience in Table~\ref{tab:m} for $n\leq 15$. 

The resummed version of the formula exactly reproduces the one-loop result
in ChPT if the tree-level scattering amplitude is inserted in the integral
in (\ref{eq:RN}), see Ref.~\cite{Colangelo:2005gd}. In the present case,
due to the special counting rules in the one-nucleon sector, the $\pi N$
amplitude receives tree-level contributions both at order $p$ as well as at
order $p^2$. Inserting the $\pi N$ amplitude up to order $p^2$ one can
indeed exactly reproduce the calculation of the $O(p^3)$ and $O(p^4)$
finite volume corrections performed in Ref.~\cite{AliKhan:2003cu}.  Since
the $\pi N$ amplitude is known well beyond the $O(p^2)$ level, in fact up
to $O(p^4)$ \cite{Becher:2001hv}, one can use this formula to go beyond the
one-loop level. In this manner the corrections beyond $O(p^4)$ will be
given exactly as far as the first three leading exponential terms are
concerned, and only approximately for the subleading exponential terms
beyond $e^{-\sqrt{3} M_\pi L}$ \cite{Colangelo:2005gd}. Based
on the analysis in Ref.~\cite{Colangelo:2006mp} we expect the contributions
which go beyond the resummed formula (\ref{eq:RN}) to be numerically small.

As seen in (\ref{eq:RN},\ref{eq:defs}), the $\pi N$ amplitude is needed
here in a particular kinematical configuration, namely for $t=0$ and for
$\nu$ purely imaginary and small: the contributions with large values of
$\nu$ are suppressed by the exponential kernel in the integral in
(\ref{eq:RN}). It is therefore natural to make a Taylor expansion of
the amplitude around $\nu=0$ after having subtracted the pole due to the
one-nucleon exchange diagram (also called the Born term). Such an expansion
is in fact already well known in the phenomenology and is referred to as
the subthreshold expansion. It reads as follows:
\be
D^+(\nu,0)=D^+_\mathrm{pv}(\nu,0)+D^+_\mathrm{p}(\nu,0)+D^+_\mathrm{na}(s,u)
\ee
where
\bea \label{eq:subt}
D^+_\mathrm{pv}(\nu,0)&=&\frac{g_{\pi N}^2}{m_N}
\frac{\nu_B^2}{\nu_B^2-\nu^2} \nonumber \\
D^+_\mathrm{p}(\nu,0)&=&d^+_{00}+d^+_{10}\nu^2+d^+_{20} \nu^4
\eea
and $\nu_B=-M^2_\pi/(2m_N)$. The function $D^+_{\mathrm{na}}(s,u)$ contains the
analytically nontrivial part of the amplitude. Up to order
$p^4$ in the chiral expansion this can be written as a sum of two single
variable functions: 
\be
D^+_\mathrm{na}(s,u)=D^+_1(s)+D^+_1(u)+O(p^5)
\ee
which admit the following dispersive representation:
\be \label{eq:disp}
D^+_1(s)=\frac{\nu^5}{\pi} \int_{M_\pi}^\infty d \nu' \frac{
  \mathrm{Im} D_1^+(s')}{\nu^{\prime 5} (\nu'-\nu -i \epsilon)}
\ee
where $s'=2m_N \nu'+m_N^2+M_\pi^2$. This representation shows that, due to
the large number of subtractions, the function $D_1^+$ is small near
$\nu=0$. As we will see later, its contribution to the finite size shift of
the nucleon mass is negligible. 

This observation leads to the following expression for the relative
finite volume shift $R_N$:
\bea\label{eq:LF}
R_N&=&\frac{3 \epp^2}{4 \pi^2} \sum_{n=1}^\infty \frac{m(n)}{\sqrt{n} \lp}
\Bigg[ g_{\pi N}^2 \epp \left(2 \pi e^{-\sqrt{n(1-\epp^2)}\lp}-\epp
    I_\mathrm{pv}(\sqrt{n} \lp, \epp)
  \right) \\
&& \hskip 1.5 cm
- \bar d^+_{00} B^0(\sqrt{n} \lp)+ \bar d^+_{10} B^2(\sqrt{n} \lp) - \bar
d^+_{20} B^4 (\sqrt{n} \lp) \Bigg] + R_{N,\mathrm{na}} \nonumber
\eea
where
\be
I_\mathrm{pv}(\lp,\epp)=\int_{-\infty}^\infty dy
\frac{e^{-\sqrt{(1+y^2)}\lp}}{\epp^2+y^2}  \; , \; \;
B^k(\lp)=\int_{-\infty}^\infty dy \; y^k e^{-\sqrt{(1+y^2)}\lp}
\label{eq:FVInt}
\ee
are the relevant finite volume integrals, $\bar d^+_{i0}=m_N M_\pi^{2i}
d^+_{i0}$ and $R_{N,\mathrm{na}}$ is the remainder coming from the
(subtracted) analytically nontrivial part of the amplitude.

If we neglect the contribution $R_{N,\mathrm{na}}$ the
representation~(\ref{eq:LF}) is very simple and expresses the finite
volume shift of the nucleon mass in terms of only a handful of physical
observables: the pion and proton masses, $M_\pi$, $m_N$, the pion-nucleon
coupling constant $g_{\pi N}$ and the three subthreshold parameters 
$\bar d^+_{i0}$. If one knows the low-energy constants which appear in the
chiral representation of these quantities, one can predict their quark mass
dependence and therefore the finite volume shift $R_N$ as a function of the
quark mass. We also stress that for simulations performed at the physical
point, Eq.~(\ref{eq:LF}) allows an evaluation of the FVC based only on
input extracted from the phenomenology, and indeed on rather well-known
quantities like $g_{\pi N}$ and the subthreshold coefficients.

\section{\texorpdfstring{Quark mass dependence of $m_N$, $g_A$, $g_{\pi N}$ and the
  subthreshold coefficients}{Quark mass dependence of mN, gA, gPiN and the
  subthreshold coefficients}}\label{sec:quarkmass}

The chiral representation for the observables $m_N$, $g_{\pi N}$ up to
order $p^4$ can be found in Refs.~\cite{Becher:1999,Becher:2001hv} and will be reproduced
here for convenience. It reads as follows:
\bea \label{eq:mgg}
 m_N &=&
m-4c_1M^2-\frac{3g^2 M^3}{32\pi F^2}+\Bigg[\tilde e_1-
\frac{3( 2 g^2 - c_2 m)}{8 N F^2} \Bigg]M^4+O(M^5) \; \; \; \; \nn
g_{\pi N}&=&\frac{g_A m_N}{F_\pi} \left(1- \frac{2 d_{18} M^2}{g}+
  O(M^4) \right) 
\eea
where $N=16 \pi^2$, the symbols $m$, $g$ and $F$ indicate the nucleon mass,
the axial coupling constant and the pion decay constant in the SU(2) chiral
limit, whereas $M^2=2 B \hat m$ is the leading term in the expansion of the
pion mass square in powers of the quark mass. The LEC $c_i$, $d_i$ and
$e_i$ appear in the chiral Lagrangian in the one-nucleon 
sector at order $p^2$, $p^3$ and $p^4$ respectively, and the tildes
indicate renormalized, scale independent versions thereof
\cite{Becher:2001hv} (the $c_i$'s as well as $d_{18}$ are scale independent
--- see Appendix \ref{app:lec} for more details). 

For the axial charge, the order $p^5$ result\footnote{We refer here to the
  counting in the Lagrangian. As seen in Eq.~(\ref{eq:gA}) this translates
  into an expression including up to the $O(M_\pi^4)$ correction in $g_A$.}
is partially known \cite{bm} and will be used in the following,
\bea\label{eq:gA}
g_A &=&
g \bigg[ 1+\left( \alpha_2 L_\chi+ \beta_2\right) \Mpi^2 + \alpha_3
  \Mpi^3 + \left( \alpha_4 L_\chi^2 + \gamma_4 L_\chi + \beta_4
  \right) \Mpi^4 \bigg]
\eea
with $L_\chi = \frac{1}{N F^2}\ln\left(\frac{\Mpi}{\mu} \right)$ and
\begin{eqnarray}
\alpha_2 \!\!&\!\!=\!\!&\!\! -2(1+2g^2)\; , \; \;
\alpha_3 = \frac{3(1+ g^2)-4 m (c_3+2 c_4)}{24 \pi F^2 m}  \; , \; \;
\alpha_4 = \frac{8}{3}+\frac{37}{3} g^2+16 g^4 \nonumber \; , \\ 
\beta_2 \!\!&\!\!=\!\!&\!\! \frac{4}{g}d_{16}^r(\mu)- \frac{g^2}{NF^2} \co
 \qquad
\beta_4 = \frac{c_4}{m}\frac{4}{N F^2} + \frac{2g^2}{(N F^2)^2}
\lbar{4} + f \co \nonumber \\
\gamma_4 \!\!&\!\!=\!\!&\!\! \frac{4(c_4-c_3)}{m}-12 d_{16}^r(\mu) \left(\frac{5}{3g}
+g \right) - \frac{2}{F^2} \alpha_2 \lbar{4} \fs
\end{eqnarray} 

Note that the above expressions for $\gamma_4$ and
$\beta_4$ do not represent the full $O(p^5)$
result. We introduce a generic LEC $f$ which collects all the polynomial
contributions proportional to $\Mpi^4$.  
The chiral expansion of the subthreshold coefficients reads: 
\bea
d_{00}^+ &=&
-\frac{2(2c_1-c_3)\mpi^2}{\fpi^2}+\frac{g^2\left(3+8g^2\right)\mpi^3}{64\pi
\fpi^4}+\nn
&&+\Mpi^4\Bigg[\frac{\tilde{e}_3}{\fpi^2}-\frac{c_1}{8 \pi^2\fpi^4}
\lbar{3}
+\frac{3\left(g^2+6g^4\right)}{64\pi^2\fpi^4m}-\frac{2c_1-c_3}{16 
  \pi^2 \fpi^4}\Bigg] \nonumber
\eea
\bea
d_{10}^+ &=&
\frac{2c_2}{\fpi^2}-\frac{\left(4+5g^4\right)\Mpi}{32\pi
  \fpi^4}+\nn
&&+\Mpi^2\Bigg[\frac{\tilde{e}_4}{\fpi^2}-\frac{16 c_1 c_2}{\fpi^2 m}-\frac{1+g^2}{4 \pi^2
  \Fpi^4 m}-\frac{197 g^4}{240 \pi^2 \fpi^4
  m}\Bigg]  \nonumber \\
d_{20}^+ &=& \frac{12+5g^4}{192 \pi \Fpi^4 \Mpi}
  +\frac{\tilde{e}_6}{\Fpi^2}+\frac{17+10g^2}{24 \pi^2 \fpi^4
  m}+\frac{173g^4}{280 \pi^2 \fpi^4 m}
\label{eq:dis}
\eea

\subsection{Determination of the LEC}
The LEC from the nucleon sector are determined by fitting the chiral
representations of $m_N$, $g_A$, $g_{\pi N}$ and $d_{i0}^+$ to experimental
data and results from lattice simulations (on $m_N$ and $g_A$
only). Details of the fits are given in Appendix \ref{app:fit}.

In order to have a better handle on the uncertainties, we perform a fit to
the order $p^2$ chiral representations as well as to the complete formulas
indicated in Eqs.~(\ref{eq:mgg}),(\ref{eq:gA}) and (\ref{eq:dis}). Note
that in the following, we give the value for the scale independent $\bar{d}_{16}$
evaluated at the physical pion mass (see Appendix \ref{app:lec})
while for the higher order LEC we indicate the values of $e_i^r(\mu), f(\mu)$ at $\mu = M_\rho$.

\begin{itemize}
\item At order $p^2$, we obtain
\begin{align}
m &= 0.896\pm0.003 \GeV &g &= 1.126\pm 0.007 \nn
c_1 &= -0.54\pm0.04 \GeV^{-1} &c_2 &= 1.79\pm0.03 \GeV^{-1} \nn
c_3 &= -3.37\pm0.18 \GeV^{-1} &\bar{d}_{16} &= 0.06 \pm 0.11 \GeV^{-2}\nn 
d_{18} &= -1.13\pm 0.19 \GeV^{-2} \nn
\chi^2/\mathrm{d.o.f.} &= 5.7/7 .
\end{align}
Note that the term proportional to $\Mpi^2$ in $g_A$ is also included
in the order $p^2$ fit, although it is a term of order $p^3$. Furthermore,
for the subthreshold coefficients, only the leading terms of the chiral
expansion are retained at order $p^2$. In particular, $d_{20}^+ = 0$.
\item The fit with the untruncated expressions of
  Sec.~\ref{sec:quarkmass} does not allow one to determine all of the
  appearing LEC. We have checked that if one holds three of the LEC at a
  constant value then the fit becomes stable. We choose to fix the three
  LEC $c_2,\; c_3$ and $c_4$. To assess the uncertainties thereby
  introduced into the analysis, we use two different sets of values for
  these three LEC, given in Ref.~\cite{Becher:2001hv},
\begin{align}
\mathrm{Set\, I}& &c_2 &= 1.7 \GeV^{-1}\co &c_3 &= -3.6 \GeV^{-1}\co &c_4 &= 2.1 \GeV^{-1}\co \nn
\mathrm{Set\, II}& &c_2 &= 2.7 \GeV^{-1}\co &c_3 &= -4.5 \GeV^{-1}\co
&c_4 &= 2.4 \GeV^{-1}\fs \nonumber
\end{align}
In the following, this fit will be referred to as the order $p^4$ fit,
although the formulas also contain terms of higher order. For set I, the fit 
yields
\begin{align}
m &= 0.907\pm0.021 \GeV &g &= 1.201\pm 0.077 \nn
c_1 &= -0.56\pm0.30 \GeV^{-1} &\bar{d}_{16} &= -3.05 \pm1.54 \GeV^{-2}\nn
d_{18} &= -1.20\pm 0.22 \GeV^{-2}  &f &= 3.2\pm 47 \GeV^{-4} \nn
e_1^r &= 15 \pm 13 \GeV^{-3} &e_3^r &= -34 \pm 72 \GeV^{-3} \nn
e_4^r &= 37 \pm 16 \GeV^{-3} &e_6^r &= 23.2\pm1.4\GeV^{-3} \nn
\chi^2/\mathrm{d.o.f.} &= 4.8/5
\end{align}
while for set II we find
\begin{align}
m &= 0.905\pm0.021 \GeV &g &= 1.207\pm 0.077 \nn
c_1 &= -0.60\pm0.30 \GeV^{-1} &\bar{d}_{16} &= -3.43 \pm 1.56\GeV^{-2} \nn
d_{18} &= -1.2 \pm 0.2 \GeV^{-2}  &f &= -0.1 \pm 47 \GeV^{-3} \nn
e_1^r &= 15 \pm 13 \GeV^{-3} &e_3^r &= 56 \pm 72 \GeV^{-3}\nn
e_4^r &= -86 \pm 20 \GeV^{-3} &e_6^r &= 23.2\pm1.4 \GeV^{-3} \nn
\chi^2/\mathrm{d.o.f.} &= 4.8 /5 .
\end{align}
\end{itemize}
The errors indicated for the LEC are statistical only and do not account,
in particular, for the uncertainties due to higher order effects. The
statistical error quoted above most likely underestimates the real
uncertainties. \newline Comparing the different resulting values of LEC,
one observes that the nucleon mass in the chiral limit lies very close to
$0.9 \GeV$ in all three fits. Also the values for $c_1$ end up quite close
together, but fall out of the error bars for $c_1$ quoted in
Ref.~\cite{GLR}. However, given the fact that the $c_i$ are correlated, we
should compare with the value of $c_1$ given in
Ref.~\cite{Becher:2001hv}, which is in better agreement. Our fits with
the two sets of order $p^2$ LEC lead to consistent results for $c_1$. The
LEC of the order $p^3$ chiral Lagrangian are less well known. Comparing
with Ref.~\cite{fm}, the values of $\bar{d}_{16}$ of the order $p^4$ fits
agree remarkably well. In the order $p^2$ fit, the resulting value of
$\bar{d}_{16}$ is not reconcilable with phenomenology (see also
Ref.~\cite{CFH}). We come to the same conclusion as the authors of
Ref.~\cite{bm}. Only the inclusion of the partial order $p^5$ corrections
in $g_A$ leads to a reasonable fit. The coupling $d_{18}$ is in agreement
with the range of values found in Ref.~\cite{fms}. Since $d_{18}$ is
determined with the help of the Goldberger-Treiman relation at the physical
point, different values of $d_{18}$ merely reflect different values of
$g$. As for the LEC of the order $p^4$ chiral Lagrangian in the nucleon
sector, the huge statistical errors of the fits confirm that they are
basically unknown.

\subsection{Numerical analysis}
\begin{figure}
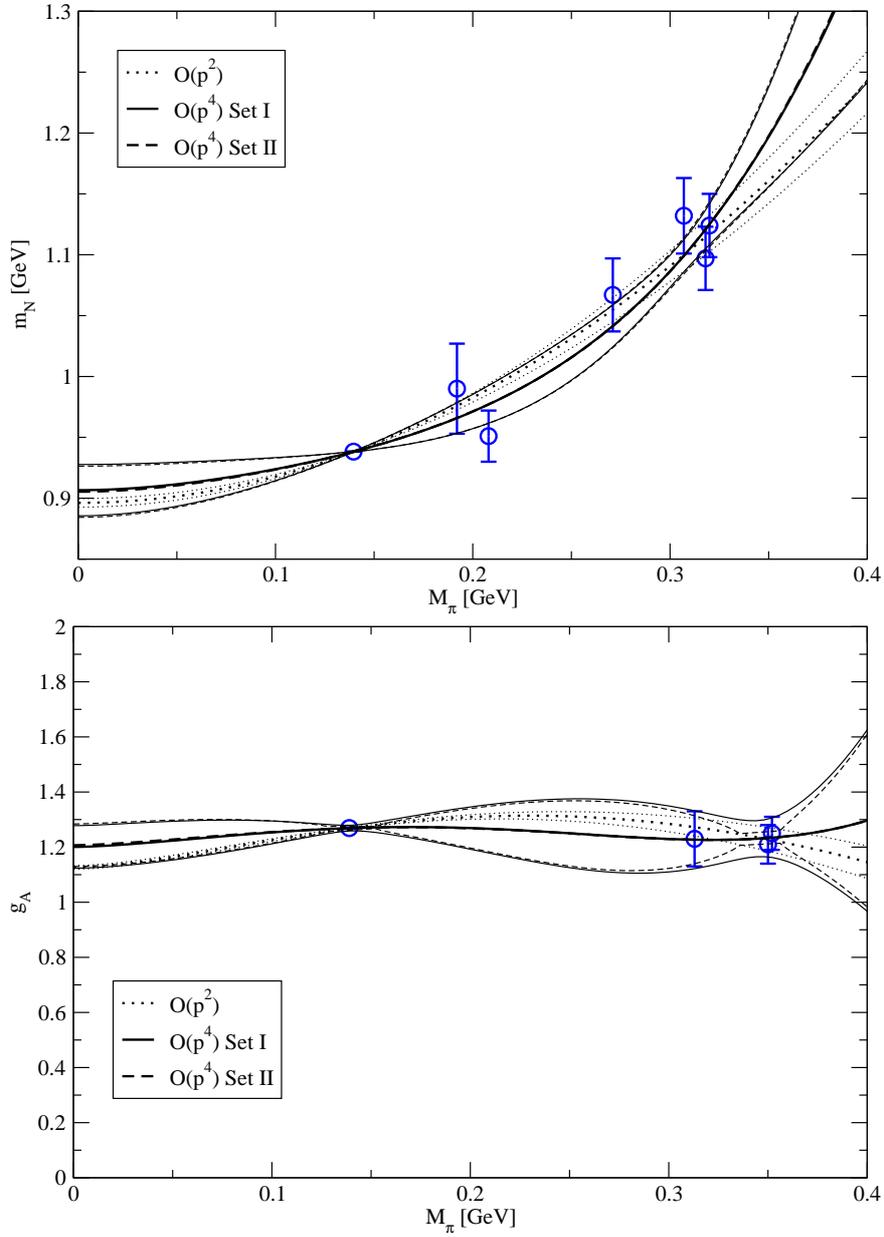

\centering
\includegraphics[width=11.5cm]{mN.eps}
\includegraphics[width=11.5cm]{gA.eps}
\caption{The upper panel shows the quark mass dependence of the
  nucleon mass and the lower panel of the axial charge, respectively. The dotted lines
  show the results of the order $p^2$ fit, the solid lines represent
  order $p^4$ set I and the dashed lines are order $p^4$ set II. The
  error bands are represented by thinner lines plotted with the same line
  style as the central value. The circles show the lattice result as
  well as the physical point used as input.}\label{fig:mNgA}
\end{figure}

\begin{figure}
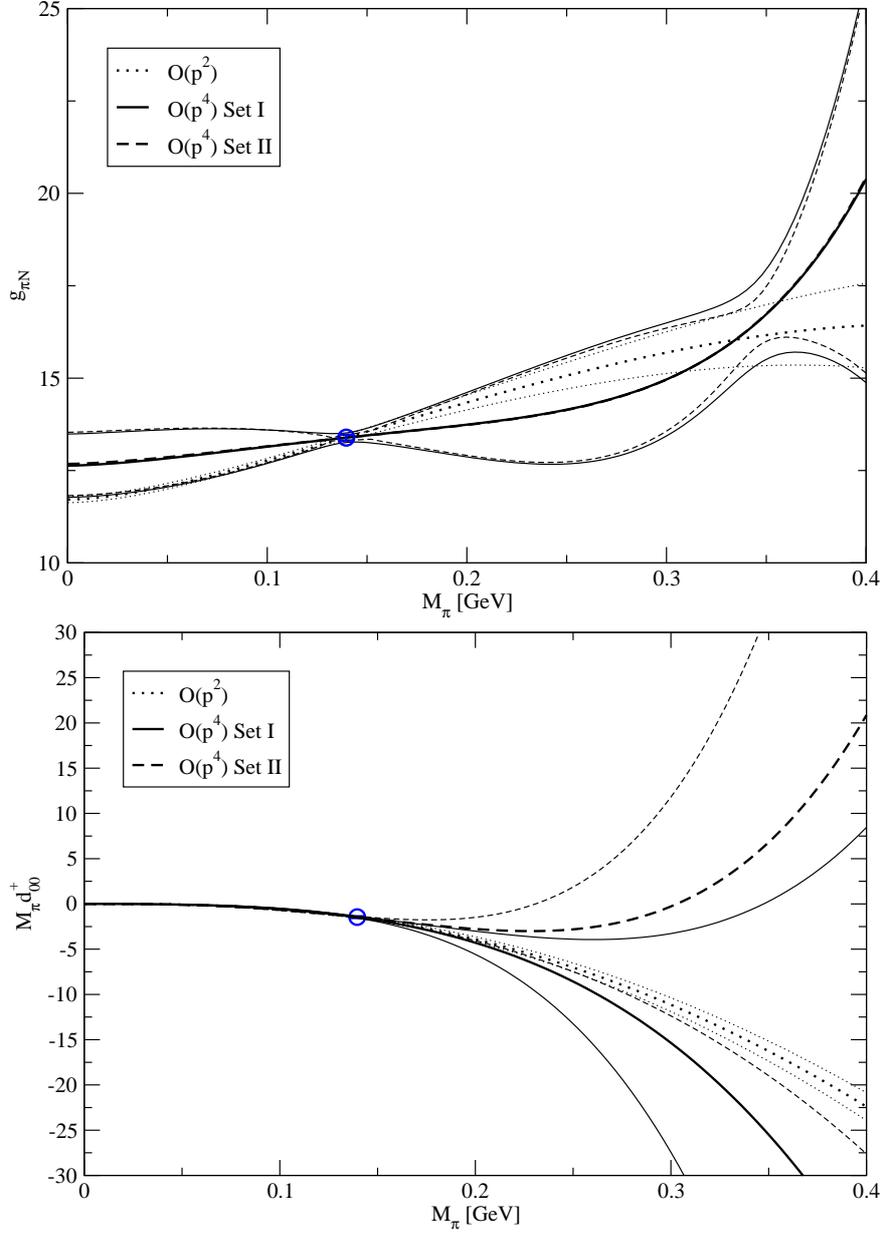

\centering
\includegraphics[width=11.5cm]{gPN.eps}
\includegraphics[width=11.5cm]{d00.eps}
\caption{The upper panel shows the quark mass dependence of the pion
  nucleon coupling constant and the lower panel of the scaled
  subthreshold coefficient $\Mpi d_{00}^+$, respectively. The dotted
  lines show the results of the order $p^2$ fit, the solid lines represent
  order $p^4$ set I and the dashed lines are order $p^4$ set II. The
  error bands are represented by thinner lines plotted with the same line
  style as the central value. The circles show the physical value used
  as input.}\label{fig:gPNd00} 
\end{figure}

Given the values of the LEC from the fits, we plot the quark mass
dependence of the quantities $m_N$, $g_A$ and the subthreshold
coefficients. As can be seen in Fig.~\ref{fig:mNgA}, the nucleon mass is
described remarkably well by the order $p^2$ curve which is just a
quadratic function in the pion mass.
Figure~\ref{fig:mNgA} also shows the axial charge. It exhibits the mild quark
mass dependence observed in lattice calculations already at order $p^2$ and
stays within the order $p^4$ error bars over a wide range of quark masses.
However, this picture hides the fact that the chiral series of $g_A$ is not
well behaved, as the observed smooth behavior is the result of
compensations among different chiral orders. We believe, however, that it is
rather unlikely to have a strong quark mass dependence in the region
between the physical pion mass and the present lowest lattice values
(around $M_\pi \sim 0.3$ GeV). As we will show in the next section the
present knowledge on the quark mass dependence of $g_A$ (which is based
mainly on lattice calculations, rather than on chiral predictions) does
allow us to calculate the FVC for the nucleon mass sufficiently reliably.
In the latter it is actually the pion nucleon coupling constant which
appears and Fig.~\ref{fig:gPNd00} shows that its quark mass dependence is
rather weak and that the difference between the two sets is negligible.

\begin{figure}
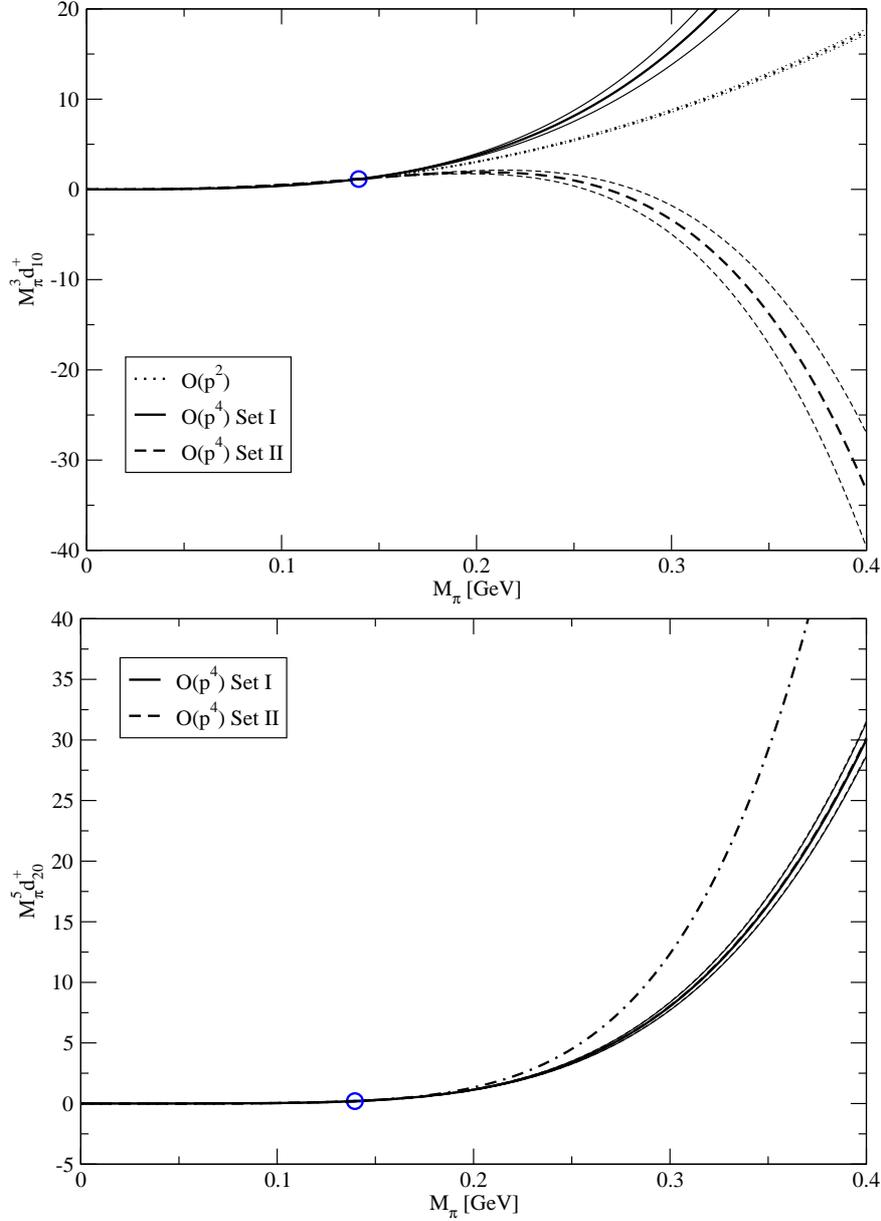

\centering
\includegraphics[width=11.5cm]{d10.eps}
\includegraphics[width=11.5cm]{d20.eps}
\caption{The upper panel shows the quark mass dependence of the scaled
  subthreshold coefficient $\Mpi^3 d_{10}^+$ and the lower panel of $\Mpi^5
  d_{20}^+$, respectively. The dotted lines show the results of the order
  $p^2$ fit, the solid lines represent order $p^4$ set I and the dashed
  lines are order $p^4$ set II. The error bands are represented by thinner
  lines plotted with the same line style as the central value. The
  circles show the physical value used as input. The dashed-dotted line
  shows the rescaled coefficient $\Mpi^5 d_{20}^+$ with $\Fpi$ replaced by
  $F$, as described in the main text.}\label{fig:d10d20}
\end{figure}

The behavior of the subthreshold coefficients $d_{00}^+$ and $d_{10}^+$
beyond the physical point is not predicted by chiral symmetry. Going from
set I to set II leads to substantial changes in the higher order LEC and
therefore to a very different behavior as a function of the quark
mass. However, the extrapolation towards the chiral limit seems to work
quite well. The coefficient $d_{20}^+$ seems to be in much better shape as
it does not depend on the $c_i$. Set I and set II almost yield identical
results. However, this nice agreement should not mislead one into
thinking that everything is under control here: large higher order
corrections may modify this nice picture, since Eq.~(\ref{eq:dis}) is the
leading order result for $d_{20}^+$. For instance, replacing $\Fpi$ by its
value in the chiral limit in $d_{20}^+$ yields the dashed-dotted curve in
Fig.~\ref{fig:d10d20}. Note that the coupling $e_6$ was readjusted to
reproduce the correct physical value.

\section{Numerical analysis of the FVC to the nucleon mass} 
\begin{figure}
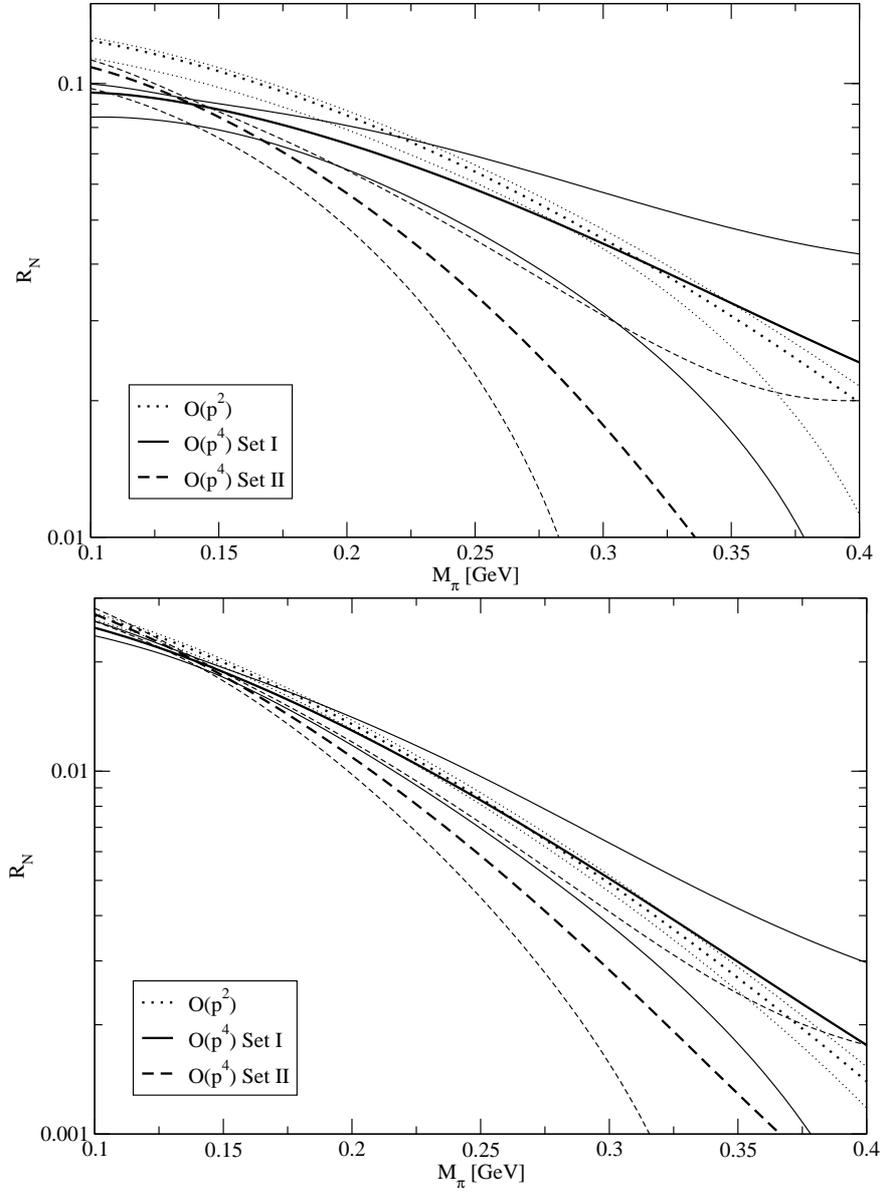

\centering
\includegraphics[width=11.3cm]{RN2fm.eps}
\includegraphics[width=11.5cm]{RN3fm.eps}
\caption{The FVC $R_N$ as a function of the quark mass for $L = 2 \fm$ (upper
  panel) and $L = 3 \fm$ (lower panel). The dotted lines indicate the result of the
order $p^2$ fit, the solid lines are the order $p^4$ fit with set I and
the dashed lines the order $p^4$ fit with set II, respectively. Note
that for $L = 2 \fm$ and $L = 3 \fm$, the pion mass should fulfill
$\Mpi \ge 0.2 \GeV$ and $\Mpi \ge 0.13 \GeV$, respectively,
in order to stay in the region of validity of the L\"uscher formula.
}\label{fig:RN2}
\end{figure}

\begin{figure}[t]
\centering
\includegraphics[width=10.5cm]{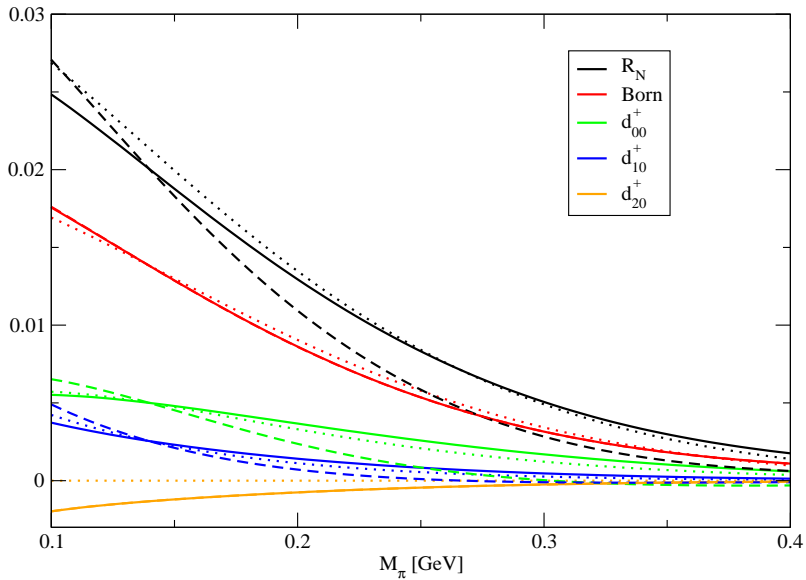}
\caption{Individual contributions to the L\"uscher formula (shown by lines
  of different colors, see legend). The finite volume effect $R_N$ (black
  lines) is the sum of all the contributions. The line style indicates the
  order and the set of LEC used. The dotted lines denote the result of the
  order $p^2$ fit, the solid lines are the order $p^4$ fit with set I and
  the dashed lines the order $p^4$ fit with set II, respectively. The size
  of the box is $L = 3 \fm$ and therefore, the pion mass should fulfill
  $\Mpi \ge 0.13 \GeV$ in order to stay in the region of validity of the
  L\"uscher formula.  }\label{fig:RNcont}
\end{figure}

With the quark mass dependence of all the quantities appearing in
Eq.~(\ref{eq:LF}) established, the finite volume effects can now be
calculated for pion masses below $\sim0.4$ GeV and box sizes $L> 2$ fm. 
In Fig.~\ref{fig:RN2} we show the finite volume effects
for boxes of spatial extent $L = 2,3 \fm$. To illustrate the size of
the different contributions to the FVC, we break down the corrections
into the different terms of Eq.~(\ref{eq:LF}). This is shown in
Fig.~\ref{fig:RNcont}. The contributions from the Born term
dominate the finite volume effects and are practically the same for all
three fits. Since the quark mass dependence of the subthreshold
coefficients is almost unknown, their contributions to the finite volume
effects vary a lot. The shift from the order $p^2$ fit to the order $p^4$
fits is mainly due to a vanishing contribution from $d_{20}^+$ at order
$p^2$. Removing this shift by setting $d_{20}^+$ to its physical value
brings the order $p^2$ curve down into the error bars of the order $p^4$
fits. \newline At order $p^4$, the fit with set I shows cancellations
between $d_{10}^+$ and $d_{20}^+$. However, these cancellations are absent
for set II, where $d_{00}^+$ and $d_{10}^+$ contribute very little but
the contribution from $d_{20}^+$ stays the same as in set I. This leads to
the large difference between set I and set II. The central values shift by
twice the error bars. It is obvious that a precise prediction of the finite
size effects is very difficult away from the physical point. One also
observes that the finite size effects strongly depend on the values of the
LEC $c_i$, $i = 2,3,4$ through the subthreshold coefficients. Therefore,
if the finite volume effects are well determined by lattice simulations,
this would provide useful constraints and possibly a determination of (some
combinations of) the $c_i$.

If we expand our formula for $R_N$ and keep in the scattering amplitude
only the $O(p^2)$ contributions we agree exactly with the one in
Ref.~\cite{AliKhan:2003cu}. A comparison of the numerics at this order is
therefore uninteresting (since by inserting the same values for the LEC one
trivially gets the same). We stress, however, that as our plots in
Fig.~\ref{fig:RN2} indicate, by truncating the chiral expansion at NLO one
may get the misleading impression of having rather moderate uncertainties
which do not grow very fast with the pion mass. The application of the
resummed L\"uscher formula allows one to go to higher orders and to verify
explicitly that:
\begin{itemize}
\item
corrections beyond NLO are important;
\item
the uncertainties grow rather quickly with the pion mass, so that going
beyond pion masses of about $0.3-0.35$ GeV they become of order 100\%.
\end{itemize}
We refrain from comparing our error analysis with Ref.~\cite{Procura:2006bj}, where
FVC including error bars were calculated at one loop in ChPT. In
Ref.~\cite{Procura:2006bj}, plots of the FVC are only given for values of pion masses
and box sizes in a region which we consider to lie outside the region of validity
of the effective theory. 

\begin{figure}[t]
\centering
\includegraphics[width=10.5cm]{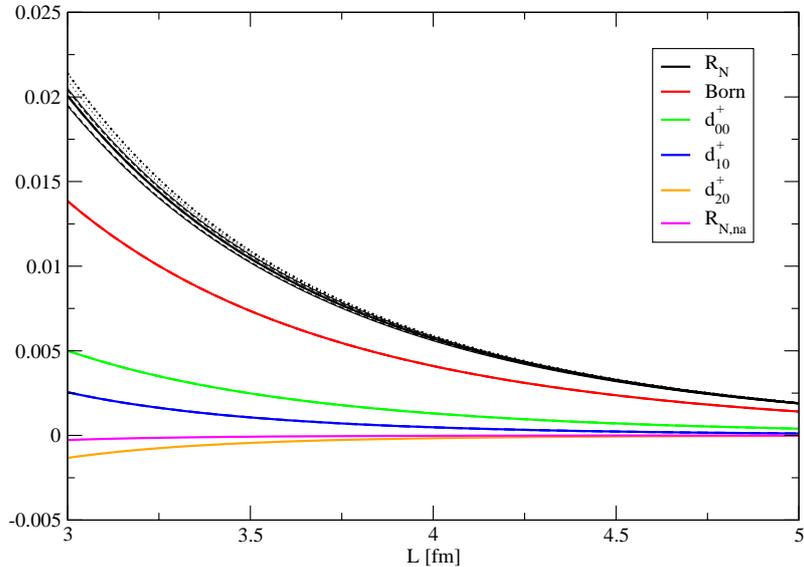}
\caption{FVC for physical pion mass as a function of the size of the box.
  The black lines denote $R_N$. The dotted line is the $O(p^2)$ fit, the
  solid line and the dashed line are set I and set II of the $O(p^4)$ fit,
  respectively. The different contributions to the corrections are also
  plotted, as well as the nonanalytic contributions $R_{N,\mathrm{na}}$
  (which are not included in $R_N$). Note that the difference between the
  order $p^2$ and order $p^4$ curves stems only from $d_{20}^+$, which
  vanishes at order $p^2$.  }\label{fig:FVCPP}
\end{figure}

Our final results are shown in Table~\ref{tab:res} where we give $R_N$ for
different values of pion masses and box sizes. The quoted central value is
the average of the central values of set I and set II and the error bars
are chosen such that they cover the error bands of both set I and set II.

\begin{table}[t]
\begin{center}
\renewcommand{\arraystretch}{1.4}
\begin{tabular*}{\textwidth}{c@{\extracolsep{\fill}}ccc}
$\Mpi [\GeV]$ & $L = 2\fm$  & $L = 2.5\fm$ & $L = 3\fm$ \\ \hline
$0.140$ & $ 0.090_{-0.006}^{+0.002\,\star}$ & $0.041_{-0.001}^{+0.001\,\star} $ &$0.0200_{-0.0005}^{+0.0004}$  \\
$0.160$ & $0.082_{-0.009}^{+0.006\,\star} $ & $0.036_{-0.003}^{+0.002}$ &$0.017_{-0.001}^{+0.001}$  \\
$0.180$ & $0.074_{-0.013}^{+0.011\,\star}$ & $0.031_{-0.004}^{+0.004}$ &$0.014_{-0.002}^{+0.002}$ \\
$0.200$ & $0.065_{-0.016}^{+0.015}$ & $0.027_{-0.005}^{+0.005}$ & $0.012_{-0.002}^{+0.002}$ \\ 
$0.220$ & $0.057_{-0.019}^{+0.019}$ & $0.023_{-0.006}^{+0.006}$  &$0.010_{-0.002}^{+0.002}$ \\
$0.240$ & $0.050_{-0.022}^{+0.022}$ & $0.019_{-0.007}^{+0.007}$ &$0.008_{-0.002}^{+0.002}$ \\
$0.260$ & $0.043_{-0.024}^{+0.024}$ & $0.016_{-0.007}^{+0.007}$  &$0.006_{-0.002}^{+0.002}$ \\
$0.280$ & $0.037_{-0.026}^{+0.025}$ &$0.013_{-0.007}^{+0.007}$ &$0.005_{-0.002}^{+0.002}$ \\
$0.300$ & $0.031_{-0.026}^{+0.026}$ &$0.011_{-0.007}^{+0.007}$ &$0.004_{-0.002}^{+0.002}$ \\
$0.320$ & $0.026_{-0.027}^{+0.027}$ & $0.009_{-0.007}^{+0.007}$ &$0.003_{-0.002}^{+0.002}$ \\
$0.340$ & $0.022_{-0.027}^{+0.027}$ & $0.007_{-0.007}^{+0.007}$  &$0.002_{-0.002}^{+0.002}$ \\
$0.360$ & $ 0.019_{-0.027}^{+0.028}$  & $0.006_{-0.007}^{+0.007}$  &$0.002_{-0.002}^{+0.002}$  \\ 
\end{tabular*}
\end{center}
\caption{Our final results for $R_N$. In
  the first column, the value of the pion mass is given and in the
  subsequent columns, the values of $R_N$ for three different sizes of
  the box are listed. The star indicates that for these values, $\Mpi
  L < 2$.}\label{tab:res}
\end{table}

\subsection{Physical point}
As we have discussed in detail in the previous sections, the main
difficulties in the numerical evaluation of the FVC for the nucleon mass
arise from the limited knowledge of the quark mass dependence of the
observables which appear in the resummed L\"uscher formula. Since several
lattice collaborations are reaching or plan to reach the physical value of
the pion mass, it is interesting to evaluate the FVC for this case ---
the advantage of our approach is that at the physical pion mass we can
insert directly the measured values for the different observables.

In Fig.~\ref{fig:FVCPP}, $R_N$ is plotted as a function of the size of the
box $L$.  The plot shows the order $p^2$ curves as well as both order $p^4$
curves. A difference is visible in $R_N$ only because at order $p^2$, the
subthreshold coefficient $d_{20}^+$ vanishes.  The main contribution to the
FVC is given by the Born amplitude. The subthreshold coefficients all yield
corrections of similar size. In fact, the contributions from $d_{10}^+$ and
$d_{20}^+$ cancel to a large extent. Since the error bands are constrained
very much by the small errors of the experimental input parameters we can
predict the FVC very accurately for lattice simulations with physical quark
masses in a box of the size $L \ge 3 \fm$.  If nucleon mass calculations
with physical quark mass are done in a box of the size $L \ge 3.5 \fm$
one does not have to worry about finite volume corrections. The 
effects stay below  1 \%.

\subsection{Systematic effects}
\label{sec:SE}
After having evaluated the size of the finite volume effects numerically,
we briefly discuss effects which have been neglected in our analysis. There
are three different types of systematic corrections:
\begin{itemize}
\item We neglect the contributions from the nonanalytic 
part of the scattering amplitude, $R_{N,\mathrm{na}}$ in
Eq.~(\ref{eq:LF}).
\item Then there are the
neglected higher order contributions of the L\"uscher formula which
vanish as $e^{-\xi \Mpi L}$, with $\xi > \sqrt{3}$.
\item Furthermore, the pole of the $\Delta(1232)$ resonance is only taken
into account as a polynomial via the LEC in the subthreshold
coefficients.
\end{itemize}
In the following we argue why it is justified to completely neglect
the first two of these effects and how we account for the last one.

We show by an explicit calculation that the nonanalytic contributions are
very small. It is possible to solve the dispersive integral Eq.~(\ref{eq:disp}) in
closed form, as described in Ref.~\cite{Becher:2001hv}. Removing the fictitious poles
from the expression 
\beq 
\hat{D}_1^+(s) = \frac{\Mpi}{12 \Fpi^4 m_N
  \nu^3}\left\{\Delta_1^++g_A^2\Delta_2^++g_A^4 \Delta_3^+\right\}
f(\nu)
\eeq with
\bea
f(\nu) &=& \frac{1}{8 \pi}
\sqrt{1-\frac{\nu^2}{\Mpi^2}}\arccos{\left(-\frac{\nu}{\Mpi}\right)},\nn
\Delta_1^+ &=& 12 \nu^4(-m_N \nu+3 \nu^2-\Mpi^2),\nn
\Delta_2^+ &=& 24 \nu^4(\nu^2-\Mpi^2),\nn
\Delta_3^+ &=& 8(\nu^2-\Mpi^2)^2(-m_N \nu +3 \nu^2+\Mpi^2),
\eea
leads to
\beq\label{dhat}
D_1^+(s) = \hat{D}_1^+(s) - \sum_{n = -3}^4 d_{1, n}^+\nu^n.
\eeq
The imaginary part of $D_1^+(s)$ vanishes identically in the L\"uscher
integral, since it is an odd function in the integration variable. 
Furthermore, in forward scattering, the contribution from $D_1^+(u)$ is the same as
the one from $D_1^+(s)$. 
Evaluating $R_{N,\mathrm{na}}$ numerically
one finds that the contributions of the nonanalytic part of the
scattering amplitude to the FVC can safely be neglected in the present
analysis. 

Higher order terms in the L\"uscher formula for finite volume
corrections to the pion mass have been explicitly
calculated in Ref.~\cite{Colangelo:2006mp}. It was found that for
$\Mpi L \gtrsim 2$, the contributions which are not included in the
resummed L\"uscher formula are very small. The formal order of
the neglected corrections to the nucleon mass is the same as for the pion
mass, $e^{-\xi \Mpi L}$. This suggests that  the relative effects of
these higher order terms for the nucleon mass are even smaller.

Finally, to estimate the effect of the $\Delta(1232)$ resonance, the particle is
included into chiral perturbation theory as an explicit degree of
freedom. The resulting tree-level scattering amplitude $D^+_\Delta$ can be found in
Ref.~\cite{Becher:1999}. Plugging this amplitude into the L\"uscher
formula Eq.~(\ref{eq:RN}) we obtain $R_\Delta$.

\begin{figure}[t]
\centering
\includegraphics[width=11.5cm]{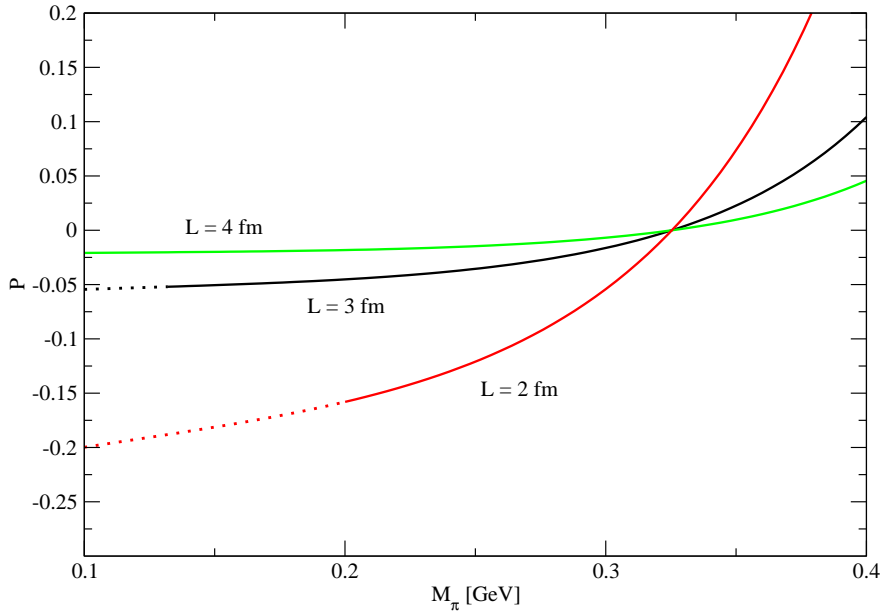}
\caption{The plot shows the ratio $P$ for different sizes of the box
  as a function of the pion mass. Note that although $P$ grows fast for larger pion masses, the absolute
  correction $R_\Delta$  vanishes exponentially. The LEC from the order
$p^4$ fit set I are used. In the region where the curves are dotted,
  $\Mpi L < 2$.}\label{fig:delta}
\end{figure}
 In Fig.~\ref{fig:delta}, we compare
$R_\Delta$ with $R_{\Delta,\mathrm{exp}}$, which is the L\"uscher
integral over the subthreshold expanded amplitude $D^+_\Delta$. The ratio $P =
(R_{\Delta,\mathrm{exp}}-R_\Delta)/R_\Delta$ is shown as a function of
the quark mass for different box sizes. The following numerical
values (from the set I fit, where relevant) are used for the nucleon
and $\Delta(1232)$ masses and the $\Delta$-nucleon coupling:  
\begin{align}\label{eq:mass}
m_N &= m \co &m_{\Delta} &= 1.232 \GeV \co &g_{\Delta N} &= 13\fs
\end{align}
Here, a comment is in
order. The additional pole of the $\Delta(1232)$ in the scattering
amplitude might lead to modifications in the L\"uscher formula, as will be
described in Sec.~\ref{sec:hm}. For our choice of masses
(see Eq.~(\ref{eq:mass})) for the nucleon and the $\Delta(1232)$, this is however not the case.

The corrections are large
because for small values of $\Mpi L$, the suppression of the
exponential factor in the L\"uscher integrand in Eq.~(\ref{eq:RN}) is
not strong enough. The region with $|\nu|/\Mpi > 1$ --- where the subthreshold
expansion does not converge anymore --- yields sizable
contributions. The absolute size of $R_\Delta$ compares to a good 
approximation with the contribution of $d_{00}^+$, but has the
opposite sign. Therefore, the use of the subthreshold expansion in the
L\"uscher formula leads to a systematic overestimation of the finite
size effects if the pole of the $\Delta(1232)$ is not included
explicitly in the $\pi N$ scattering amplitude. We account for this
effect by introducing an asymmetric final error $R_N \pm \sigma^\pm_{R_N}$
for the finite size effects, 
\begin{align}
\sigma^+_{R_N} &= \sigma_{R_N} \co &\sigma^-_{R_N} &= \sqrt{\sigma_{R_N}^2+(P
    R_\Delta)^2}\fs
\end{align}
The definition of $\sigma_{R_N}$ is provided in Appendix \ref{app:fit}.
\section{Resummed L\"uscher formula for the heavy meson masses}\label{sec:hm}

In this section we present a formula that we will use for analyzing the
finite volume corrections to the mass of a heavy meson similar to the one
in Eq.~\eqref{eq:RN}. Again we use a resummed version of the L\"uscher formula.
In the literature one finds a relativistic \cite{Burdman:1992gh} as well as a
nonrelativistic \cite{Wise:1992hn} effective
Lagrangian that can be used to calculate the scattering amplitude. They are very
briefly introduced in Appendix~\ref{sec:HMChPT}, where we also give further
references.


Indeed for both formalisms we can provide a L\"uscher formula and then calculate
the finite volume mass shift. We first discuss the formula that takes the
relativistic amplitude as an input. The expression for the relative finite
volume correction $R_\HM^\mathrm{rel}\equiv(\mH(L)-\mH)/\mH$ in this case reads as follows 
\begin{multline}
	R_H^\mathrm{rel} = \frac{\mpi^2 }{32 \pi^2 \mH^2} \sum_{n=1}^{\infty}
	\frac{m(n)}{\sqrt{n} \lp }\left[ 2 \pi \eppb \gHM^2
	e^{-\sqrt{n(1-\eppb^2)}\lp}\; \theta(\mpi^2 + \mH^2 - \mHs^2)
	\phantom{\int}\right. \\[3mm]
	- \left. \int_{-\infty}^{\infty} \! dy
	e^{-\sqrt{n(1+y^2)}\lp} \tilde T^{+}(y) \right] \;,
	\label{eq:RB}
\end{multline}
where
\begin{equation}
	\eppb = \frac{\mpi^2 + \mH^2 - \mHs^2}{2 \mH \mpi}\; , 
	\; \; \lp= \mpi L \; , \; \; \tilde T^{+}(y) = T^+(i \mpi y,0) .
\end{equation}
Here and later we adopt the notation that \HM\ and \HMs\ stand for any of
the heavy pseudoscalar and vector mesons, respectively.
The L\"uscher formula is valid provided that
\begin{equation}
	\Delta_{*} \equiv \mHs - \mH \geq 0, \qquad \mpi, \Delta_{*} \ll \mH, \mHs,
	\qquad \lp \gtrsim 2.
	\label{eq:RBrestr}
\end{equation}
The first two relations are well satisfied for all physical masses,
while the third one requires $L \gtrsim 3 \fm$ at the physical pion mass.
$T^+$ is part of the elastic $\pi \HM$ scattering amplitude, defined as
\begin{equation}
	T(\pi^a(q) \HM(p) \to \pi^{a'}(q') \HM(p')) \equiv
	T_{a'a} = \delta_{a'a} T^+ + \frac{1}{2} [\tau_{a'},\tau_a] T^- \; ,
\end{equation}
where the amplitudes $T^\pm$ depend on $\nu$ and $t$, which are defined similarly
as in Eq.~(\ref{eq:nutN}):
\begin{equation}
	 t=(q-q')^2 \; ,\; \; \nu=\frac{s-u}{4 \mH} \; , \quad \mbox{where} \quad 
	s=(p+q)^2\; , \; \; \; \; u=(p-q')^2 \; .
\end{equation}

The proof of this formula is done in the same spirit as in the original
article by L\"uscher \cite{Luscher:1985dn}.  The crucial difference is that
the interaction of a heavy pseudoscalar meson \HM\ with a pion is
mediated by the heavy vector meson \HMs, which has a slightly larger mass than
the \HM. At low energy the pole generated by the \HMs exchange is the most
important singularity and the one which dominates the L\"uscher formula. 
If one does not take the $m_q \to \infty$ limit and considers \HMs\ as
nondegenerate with \HM, a third mass enters the FVC analysis and leads to
some modifications of the proof by L\"uscher: the pole that 
is responsible for the first term in Eq.~\eqref{eq:RB} is shifted outside
of the integration contour for some configurations of the meson masses such
that its contribution to the mass shift vanishes (see Fig.~\ref{fig:nu}). This
is accounted for by the step function.
Note that the exponential in the first term can even become complex for some
values of the masses, as the constraints given in Eq.~\eqref{eq:RBrestr}
do not restrict $\eppb$ to values smaller than 1. However, this only occurs for
mass values where the step function vanishes such that the resulting mass shift
remains real.

We stress that in the degenerate case $\mH = \mHs$, Eq.~\eqref{eq:RB} takes
the same form (modulo trivial overall factors) as the one for the nucleon
mass --- which is the reason for discussing the two in the same paper. What
makes this case particularly interesting is precisely the presence of a
second almost degenerate mass, whose exact value we are free to vary. By
varying it, we can better study and understand the role of the first,
exponentially leading term in the L\"uscher formula, and ask, in particular,
what happens if we change the masses such that the step function
vanishes, and the first term in Eq.~(\ref{eq:RB}) disappears. This question
is also relevant for the nucleon case if we consider the contribution of
the $\Delta$, which we have briefly discussed in Sec.~\ref{sec:SE}. Indeed
as we vary the pion mass, the $\Delta$ mass may get closer (depending on
its quark mass dependence) to the nucleon mass and the corresponding pole
may move from the right-hand to the left-hand side of the imaginary axis in
the $\nu$-plane. It is then a relevant question, what influence this has on
the FVC. The dependence of the FVC on the mass difference $\Delta_{*}$ will
be discussed below.

The L\"uscher formula takes as input the amplitude in forward direction, i.e.
for $t=0$, and for $\nu$ purely imaginary and small, as the contributions for
large $\nu$ are suppressed due to the exponential in the integrand. Similarly as
for the $\pi N$ amplitude, it is convenient to use the subthreshold expansion
of the $\pi H$ amplitude:
\begin{equation}
	T^+(\nu,0)=T^+_\mathrm{pv}(\nu,0)+T^+_\mathrm{p}(\nu,0)+T^+_\mathrm{na}(s,u)\;,
\end{equation}
where
\begin{equation}
	\begin{split}
		T^+_\mathrm{pv}(\nu,0) &= \frac{\gHM^2 \nu_C^2}{\nu_C^2-\nu^2} \;,
\\[2mm]
		T^+_\mathrm{p}(\nu,0) &= t^+_{00}+t^+_{10}\nu^2+t^+_{20} \nu^4 \; ,
	\end{split}
\end{equation}
with
\begin{equation}
	\nu_C = -\eppb \mpi = -\frac{\Mpi^2 + \mH^2 - \mHs^2}{2 \mH}\; . 
\end{equation}
In contrast to the subthreshold expansion for the $\pi N$ amplitude, there is no
inverse mass factor in the definition of $T^+_\mathrm{pv}$, because here the amplitude
is dimensionless. In order to have a dimensionless coupling constant \gHM, 
we also need to apply a definition that is somewhat different from the usual one, namely
\begin{equation}
	\gHM^2 = \lim_{\nu^2 \rightarrow \nu_C^2} \;
\frac{\nu_C^2-\nu^2}{\nu_C^2} \; T \;.
\end{equation}

Inserting the subthreshold expansion in the formula for the mass shift in 
Eq.~(\ref{eq:RB}), we get
\begin{multline}
	R_\HM^\mathrm{rel} = \frac{\mpi^2 }{32 \pi^2 \mH^2} \sum_{n=1}^{\infty}
	\frac{m(n)}{\sqrt{n} \lp } \\[3mm]
	\times \left[ \gHM^2 \eppb
	\left( 2 \pi e^{-\sqrt{n(1-\eppb^2)}\lp}\; \theta(\mpi^2 + \mH^2 - \mHs^2)
	- \eppb I_\mathrm{pv} (\sqrt{n} \lp,\eppb ) \right) \right.\\[3mm]
	\left. \phantom{e^{\sqrt{n}}} - \bar t_{00}^+ B^0(\sqrt{n} \lp) + \bar t_{10}^+ B^2(\sqrt{n} \lp)
	-\bar t_{20}^+ B^4(\sqrt{n} \lp) \right] + R_{H,\mathrm{na}} \;.
	\label{eq:RBb}
\end{multline}
The finite volume integrals $I_\mathrm{pv}$ and $B^k$ are the same as in
Eq.~\eqref{eq:FVInt}. Furthermore, $\bar t^+_{i0} = \mpi^{2i}
t^+_{i0}$ and $R_{\HM,\mathrm{na}}$ is the remainder coming from the
analytically nontrivial part of the amplitude.

At tree level, we find for the subthreshold coefficients
\begin{align}
		t^+_{00} &= \frac{\gHM^2 (\mH \nu_C + 2 \mpi^2) \nu_C}
		{\mH (\mpi^2 - \nu_C^2)},
		&t^+_{10} &= t^+_{20} = 0  \; ,
\end{align}
with the coupling constant given by
\begin{equation}
	\gHM^2 = \frac{3 g^2 \mH^3}{F_\pi^2 \mHs^2 \nu_C} 
	( \nu_C^2 - \mpi^2 ) \; .
\end{equation}
Furthermore, $T^+_\mathrm{na}(s,u) = 0$,
as no nonanalytical contributions can come from tree-level graphs, which
then implies that $R_{\HM,\mathrm{na}} = 0$.


The expression for the relative finite volume correction
$R_\HM^\mathrm{nrel} \equiv (\mH(L)-\mH)/\mH$ 
in the nonrelativistic formalism reads
\begin{equation}
	R_\HM^\mathrm{nrel} = -\frac{\mpi^2 }{32 \pi^2 \mH} \sum_{n=1}^{\infty}
	\frac{m(n)}{\sqrt{n} \lp } \int_{-\infty}^{\infty} \! dy
	e^{-\sqrt{n(1+y^2)}\lp} \tilde T^{+}(y) \;.
	\label{eq:RBnrel}
\end{equation}
Note that the absolute mass difference $\mH R_\HM^\mathrm{nrel}$ is now
independent of \mH. The term proportional to $\gHM^2$ has
disappeared because the argument of the step function is always negative if we
take the limit $m_H \to \infty$ at fixed $\Delta_*$ (no matter how
small it is). 

The proof of the L\"uscher formula only depends on the positions of the poles
of the propagators, which are the same also in the nonrelativistic theory.
Thus there is no need to repeat the full proof, all we have to do is to expand the
absolute mass difference $\mH R_\HM^\mathrm{nrel}$ in Eq.~\eqref{eq:RB} to
leading order in $1/\mH$. As pointed out in
Appendix~\ref{sec:HMChPT}, the fields used for the two Lagrangians differ in
normalization and to compensate for this we have to absorb a factor $1/\mH$ into
the scattering amplitude $\tilde T^{+}$. 

Because the nonrelativistic amplitude has mass dimension $-1$, we have again to change
the definition of the coupling \gHM\ to
\begin{equation}
	\gHM^2 = \lim_{\nu^2 \rightarrow \Delta_*^2} \;
\frac{\Delta_*^2-\nu^2}{\Delta_*} \; T \;,
\end{equation}
which is dimensionless. This redefinition then leads to
\begin{equation}
	T^+_\mathrm{pv}(\nu,0) = \frac{\gHM^2 \Delta_* }{\Delta_*^2-\nu^2}.
\end{equation}

Inserting the subthreshold expansion
in the formula for the finite volume effect in Eq.~\eqref{eq:RBnrel}, we get
\begin{multline}
	R_\HM^\mathrm{nrel} = \frac{\Mpi^2}{32 \pi^2 \mH} \sum_{n=1}^{\infty}
	\frac{m(n)}{\sqrt{n} \lp }
	\left[ -\frac{\gHM^2}{\Delta_*}\, \eppt^2 I_\mathrm{pv} (\sqrt{n} \lp, \eppt) \right. \\[3mm]
	\left. \phantom{\frac{1}{1}} - \bar t_{00}^+ B^0(\sqrt{n} \lp) + \bar t_{10}^+ B^2(\sqrt{n} \lp)
	-\bar t_{20}^+ B^4(\sqrt{n} \lp) \right] + R_{\HM,\mathrm{na}} \; .
	\label{eq:RBnrelb}
\end{multline}
The finite volume integrals $I_\mathrm{pv}$ and $B^k$ are again the same as in
Eq.~\eqref{eq:FVInt} and 
\begin{equation}
	\eppt = -\frac{\Delta_*}{\mpi} \;.
\end{equation}
Also, $\bar t^+_{i0}=\mpi^{2i}t^+_{i0}$ and $R_{\HM,\mathrm{na}}$ is the remainder
coming from the analytically nontrivial part of the amplitude.

From the nonrelativistic Lagrangian we then get at tree level,
\begin{align}
	t^+_{00} &= \frac{\gHM^2 \Delta_*} {\mpi^2 - \Delta_*^2}, 
	&t^+_{10} = t^+_{20} &= 0 \; , \\[2.5mm]
	\gHM^2 &= \frac{3 g^2}{F_\pi^2}(\Delta_*^2-\mpi^2) \; ,
	&T^+_\mathrm{na}(s,u) &= 0.
\end{align}
These results are identical to the leading term in the $1/\mH$ expansion of
the corresponding relativistic expressions (up to factors of $\mH$ and
$\Delta_*$  due to the normalization of the heavy meson field and
the differing definition of \gHM, respectively).

\section{Numerical analysis of the FVC to the heavy meson masses}
The finite volume effects can now be calculated for arbitrary pion and
heavy meson masses and box sizes $L$ within the validity of chiral perturbation
theory. To get meaningful results, one has to
ensure that the conditions given in Eq.~\eqref{eq:RBrestr} are respected.

For the coupling constant $g$ a number of values from different sources are
available in the literature (see Refs.~\cite{Khodjamirian:1999hb,
  Anastassov:2001cw, Abada:2002xe, Abada:2002vj, Abada:2003un,
  Becirevic:2005zu, Becirevic:2009yb}). $g$ depends on both the mass of
the heavy and the light quark, and neither dependence is fully
understood. The listed publications give values for $g$ for different heavy
quark and light quark masses and, including also the error bars, these
range from 0.18 to 0.79. For the numerical results presented in the
following, we have used $g = 0.5$ everywhere and thus neglected any quark mass
dependence.

\begin{figure}[p]
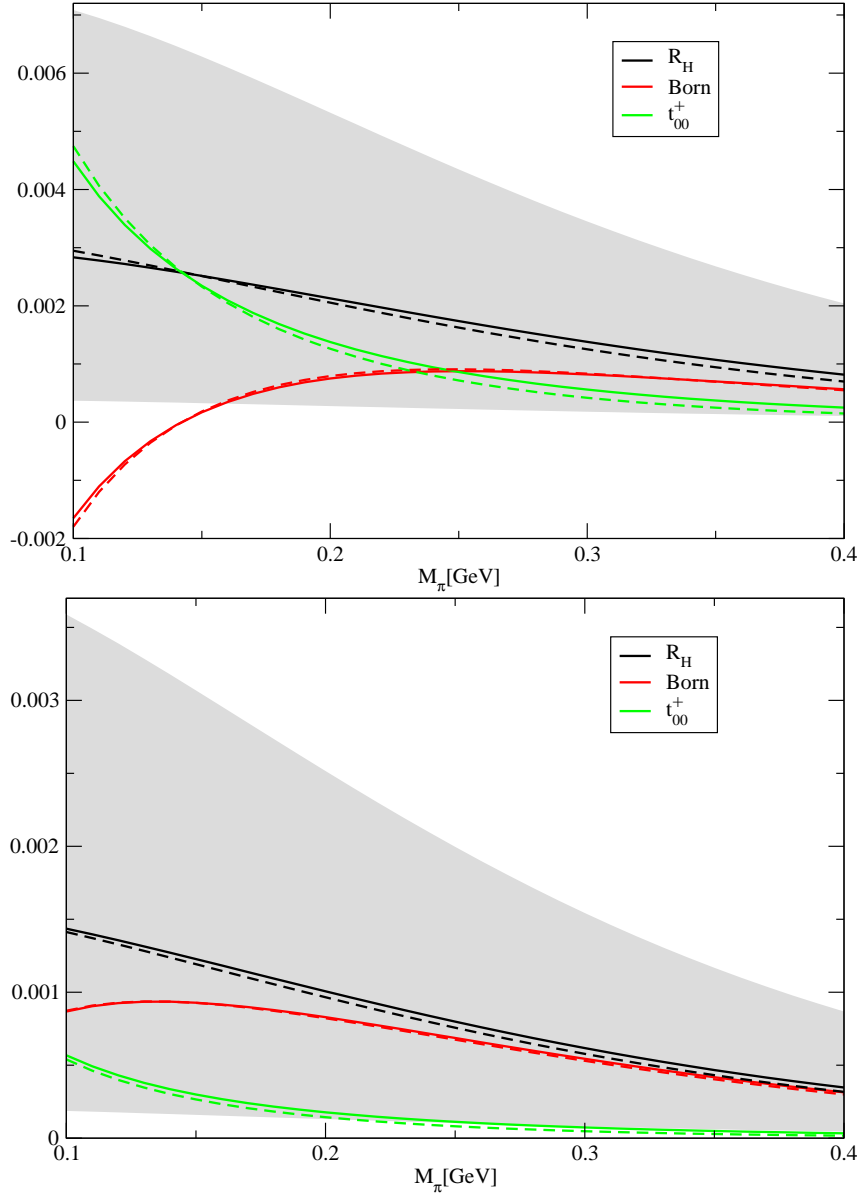

\centering
\includegraphics[width=11.2cm]{RBD2fm.eps}
\includegraphics[width=11.2cm]{RBB2fm.eps}
\caption{Finite volume effects as a function of the pion mass for a box
  size of $L = 2 \fm$, in the upper panel for the $D$, in the lower panel
  for the $B$ meson. The contributions of the different terms to the
  L\"uscher formula are shown separately, the finite volume effect $R_\HM$
  is given by the sum. Solid lines are for the relativistic, dashed lines
  for the nonrelativistic formalism, the shaded area is the uncertainty on
  the relativistic result coming from the coupling constant $g$. Note that
  for $L = 2 \fm$, the pion mass should fulfill $\Mpi \ge 0.2 \GeV$ in
  order to stay in the region of validity of the L\"uscher formula.
}\label{fig:RB2}
\end{figure}

\begin{figure}[p]
\centering
\includegraphics[width=11.2cm]{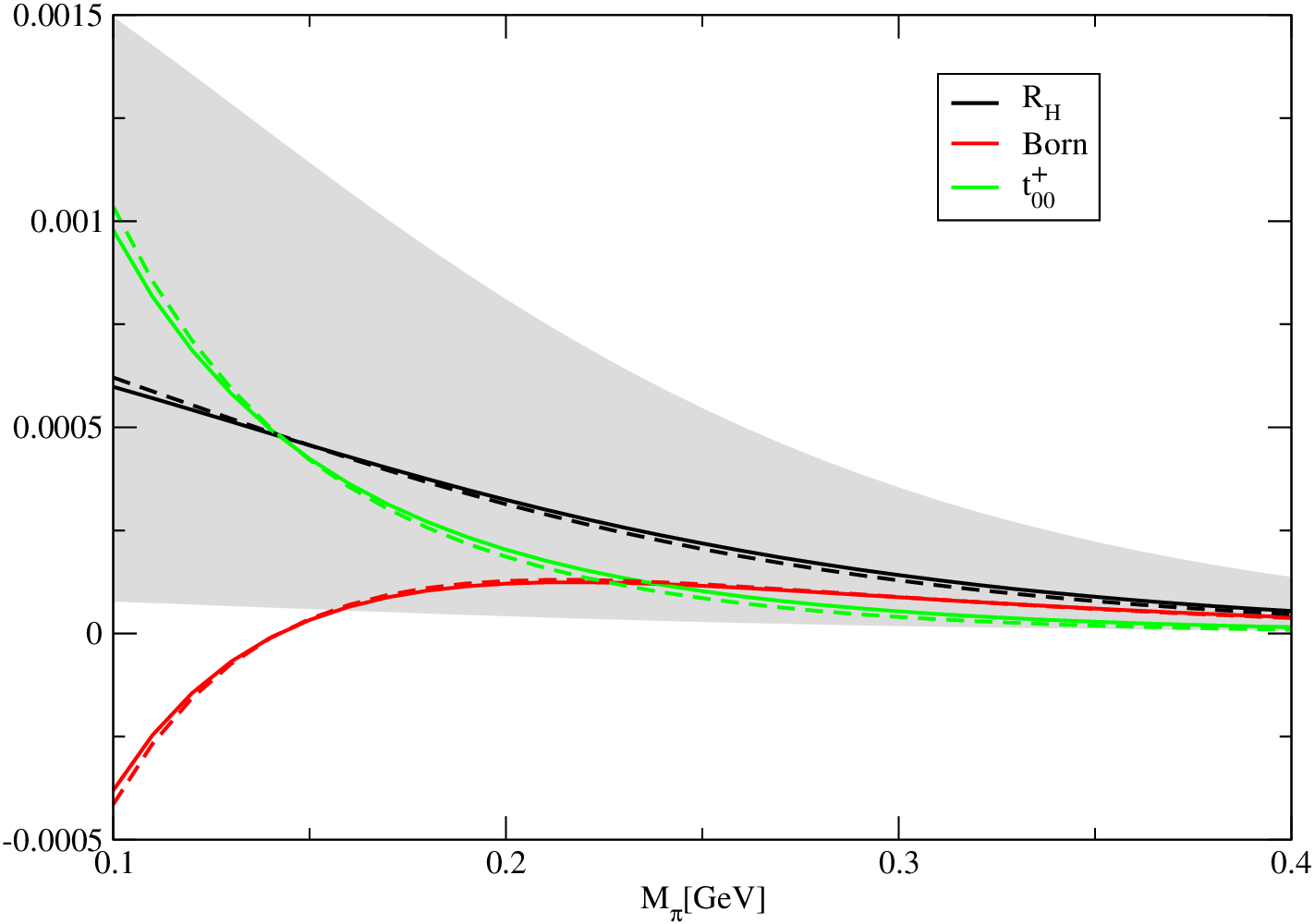}
\includegraphics[width=11.2cm]{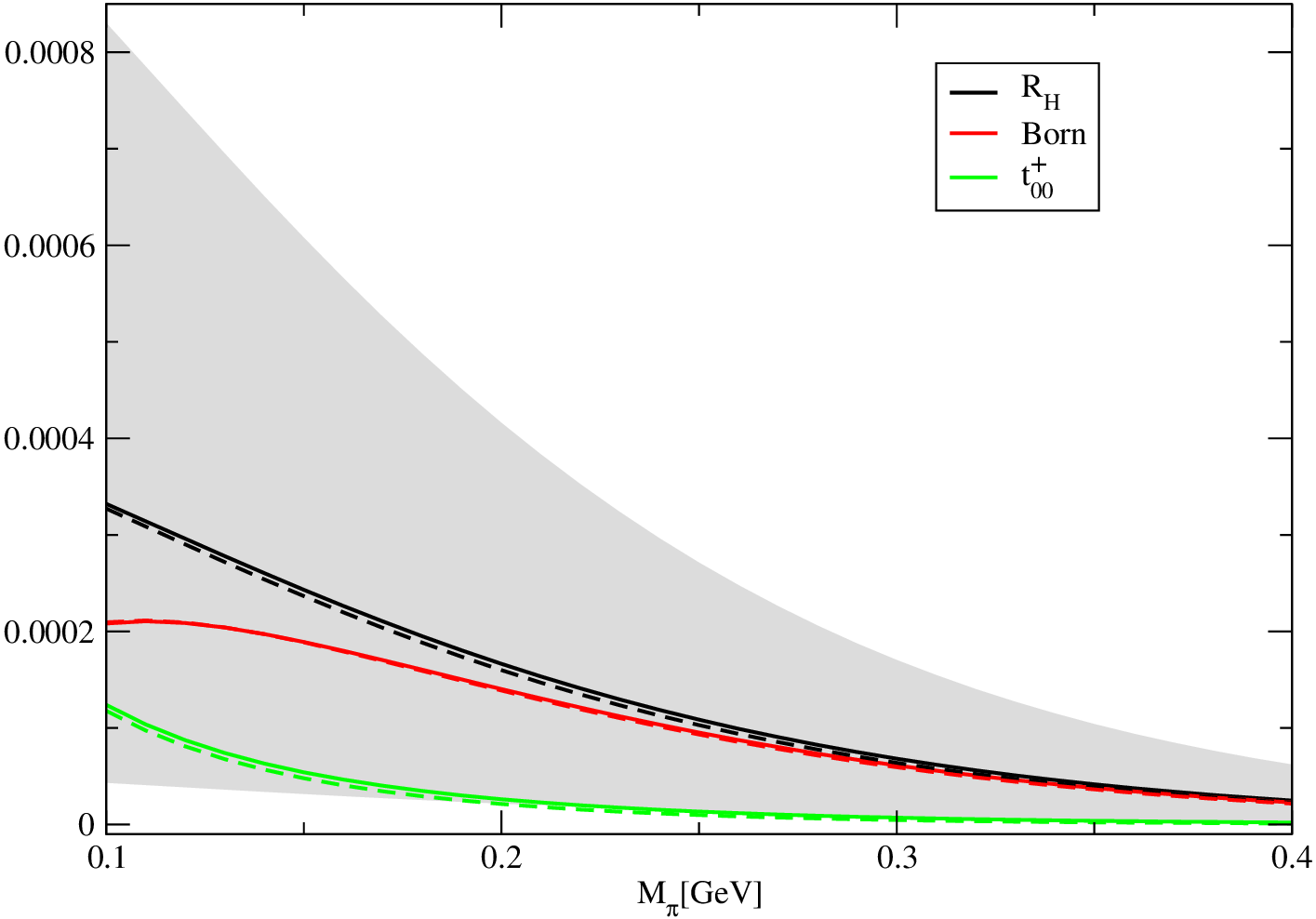}
\caption{Finite volume effects as a function of the pion mass for a box
  size of $L = 3 \fm$: upper panel for the $D$, lower panel for the $B$
  meson. The contributions of the different terms to the L\"uscher formula
  are shown separately, the finite volume effect $R_\HM$ is given by the
  sum. Solid lines are for the relativistic, dashed lines for the
  nonrelativistic formalism, the shaded area is the uncertainty on the
  relativistic result coming from the coupling constant $g$.  Note that for
  $L = 3 \fm$, the pion mass should fulfill $\Mpi \ge 0.13 \GeV$ in order
  to stay in the region of validity of the L\"uscher formula.
}\label{fig:RB3}
\end{figure}

Figures~\ref{fig:RB2} and \ref{fig:RB3} show the finite volume effects
for the $D$ and $B$ meson as a function of the pion mass for box sizes
$L = 2, 3 \fm$. The contributions coming from the Born term and from $t_{00}^+$ are 
plotted separately for both, the relativistic and the nonrelativistic L\"uscher
formula. The range of values covered by varying $g$ from
$0.18$ to $0.79$  in the relativistic formula is shaded in gray. Despite
the large uncertainties the main message to be taken from the figures is
that these FVC are negligibly small even for volumes as small as $L=2$
fm. 

We find it nonetheless instructive to discuss in some detail some features
which one can read off from the numerical analysis. First of all we remark
that the deviation between the relativistic and the nonrelativistic result comes
mainly from the $t_{00}^+$ term and it is very small compared to the uncertainty coming
from the coupling constant. A detailed analysis of the size of the deviation
will be given below.

For the $B$ meson, the Born term compensates parts of the very large $t_{00}^+$
contribution at small pion masses, while for larger pion masses, both contributions
are positive and of similar size. For the $D$ meson the Born term clearly dominates
the finite volume effect. The net effect is smaller for the heavier $B$ meson and
the $t_{00}^+$ contribution is reduced more than the Born term, as it is suppressed
by an additional factor of $1/\mH$.

In order to compare the results from the relativistic and the nonrelativistic 
framework we define the relative difference between the two as
\begin{equation}
	\delta_\HM = R_\HM^\mathrm{rel}/R_\HM^\mathrm{nrel} - 1 \;.
\end{equation}
Numerical evaluation shows that $\delta_\HM$ indeed goes to zero for
very large heavy meson masses. In the following, we give $\delta_\HM$ for
physical values of the masses for a box size of $L = 3 \fm$. For the $D$ meson
with $m_D = 1.87 \GeV$ and $\Delta_* = 0.142 \GeV$ it is about 0.75\%,
for the $B$ meson with $m_B = 5.28 \GeV$ and $\Delta_* = 0.046 \GeV$
about 2.5\%.  Because $R_\HM$ is proportional to $g^2$, $\delta_\HM$ is
independent of the coupling constant and consequently we do not give an error
on these numbers. The absolute value of $\delta_\HM$ is very slowly decreasing
with growing $L$ such that the given values are good estimates for any box size.
For unphysical pion masses, $\delta_\HM$ can become quite large, indicating that
the nonrelativistic framework is not very well suited for large pion masses.
For $\Mpi = 0.4 \GeV$, say, it is about 10\% for the $B$ and about 17\% for the
$D$ meson.

Surprisingly, despite being the leading order term of a $1/\mH$ expansion,
the nonrelativistic framework is more accurate for the lighter $D$ meson.
The reason is that $\delta_H$ also depends on the mass splitting
$\Delta_*$. If no heavy quark symmetry breaking effects are taken into
account, i.e. when $\Delta_* = 0$, the agreement is better for the $B$
meson.

\begin{figure}[t]
\centering
\includegraphics[width=11.5cm]{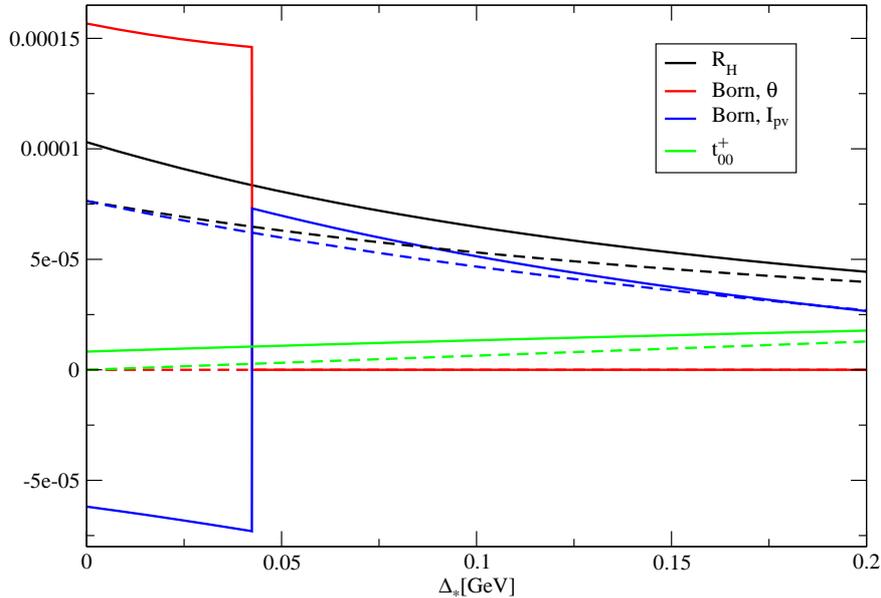}
\caption{Finite volume effect for the $D$ meson as a function
of $\Delta_*$ with $L = 3 \fm$ and $\mpi = 0.4 \GeV$. Solid lines
are for the relativistic, dashed lines for the nonrelativistic formalism.
The Born term has been split up in the contribution containing the step function
and the one with the integral $I_{\mbox{pv}}$. The latter has a discontinuity,
which is compensated by the former that starts to contribute exactly at this
point. The error band has been omitted.
}\label{fig:RBDelta}
\end{figure}
We add some comments on the role of the step function in
Eq.~\eqref{eq:RBb}, which is nonzero if
\begin{equation}
\Delta_* < \tilde \Delta_* = \sqrt{\mH^2 + \mpi^2} - \mH \approx
\frac{\Mpi^2}{2 \mH}\;. 
\end{equation}
By varying the parameter $\Delta_*$ between zero and the physical value
observed for the case of the $D$ and $B$ mesons we can interpolate between
the case in which the singularity structure of the amplitude is exactly
like in the case of the nucleon (formally reached in the infinite quark
mass limit), and the physical case with the \HMs\ meson nondengenerate with
the \HM\ meson. This corresponds to moving the pole in the $\nu$-plane from
the left-hand side to the right-hand side of the imaginary axis, as shown
in Fig.~\ref{fig:nu}. How this affects the FVC is demonstrated in
Fig.~\ref{fig:RBDelta}, where the finite volume effect is plotted as a function
of the mass splitting $\Delta_*$. There one can clearly see the point where the
step function starts to contribute. At the same point, the other contribution to
the Born term, containing the integral $I_\mathrm{pv}$, is
discontinuous. This is due to the fact that $\eppb$ goes through zero at
$\Delta_* = \tilde \Delta_*$, while the product $\eppb I_\mathrm{pv}$ goes to a
nonzero value in the limit $\eppb \to 0$ and $I_\mathrm{pv}$ is even in
$\eppb$. 
The term proportional to the step function compensates this effect
such that the resulting finite volume effect is continuous. Indeed this term
arises from a deformation of the
integration path which keeps the pole due to \HMs\ particle always to its
right-hand side of the path, even when the pole moves to the left-hand side of the
imaginary axis. Since in this way the integration path does not cross any singularity
as we change $\Delta_*$, the integral remains analytic in $\Delta_*$ and we
get a continuous curve. The figure also shows that in the nonrelativistic
version of the L\"uscher formula the mass shift is continuous even without
the step function term, because $\eppt$ is always negative for $\Delta_*
\ge 0$ and thus the point of the sign change is not in the range of
validity of the L\"uscher formula.

Other than for the nucleon, where the physical observables entering the
L\"uscher formula are best known at physical quark mass, the limited
knowledge on the coupling constant \gHM\ does not allow for a better
accuracy on the FVC at the physical point in the case of the heavy meson. Also,
it is very challenging to perform lattice simulations for heavy mesons with
very small masses for the light quarks and thus we refrain from discussing this
situation in detail.

\section{Summary and conclusions}
In this paper we have evaluated the finite volume corrections to the mass
of nucleons and heavy mesons in the $p$-regime on the basis of the resummed
L\"uscher formula. The latter relates the finite size effect for the mass
of a particle $P$ to an integral over the forward $\pi P$ scattering
amplitude. We give simple analytic formulas which express the finite
volume corrections to the nucleon and heavy meson masses in terms of only a
handful of physical observables.

In the case of the nucleon, these observables are the pion mass
$\Mpi$, the proton mass $m_N$, the pion--nucleon coupling constant
$g_{\pi N}$ and the subthreshold parameters $d_{i0}^+$. Relying on the
chiral representation of these quantities, one only needs to know the
values of the chiral low-energy constants (LEC) to be able to predict the
finite volume effects as a function of the quark mass.
Fitting the nucleon mass $m_N$ and the axial charge $g_A$ to lattice
results as well as fixing the other observables at the physical point,
we extract values for these LEC. These values are then used as a
basis for our numerical analysis which shows that:
\begin{itemize}
\item At the physical point, i.e. for a pion mass of $\Mpi = 0.140
  \GeV$, the relative finite size effects drop
  below $1 \%$ for $L \geq 3.5 \fm$. 
\item At the physical point and for box sizes of $L \geq 3\fm $  we can predict
  the relative finite size effect with a precision of about 1\textperthousand.
\item For unphysical pion masses, the relative finite size effects stay
  below $10 \%$. However, moving away from the physical point, the
  precision deteriorates quickly and an accurate prediction of the finite
  size effect becomes impossible. For instance, in a box of the size $L =
  2\fm$ and for $\Mpi = 0.3\GeV$, the relative finite size effect can be
  between $0 \%$ and $6 \%$.
\item The finite size effects strongly depend on the order $p^2$ LEC
  $c_i$ and on the quark mass dependence of $g_{\pi N}$ (and therefore of
  $g_A$). Future analyses of better lattice data should exploit these
  relations to extract interesting information about these quantities.
\item Our final results for the relative FVC are tabulated in
  Table~\ref{tab:res}. 
\end{itemize}
We stress that it is thanks to the resummed L\"uscher formula that one can
relatively easily evaluate higher order contributions to the final volume
effects and better assess the uncertainties. At the plain one-loop level in
chiral perturbation theory \cite{AliKhan:2003cu}, for example, the quark
mass dependence of $g_A$ does not yet play a role, and the contribution of
the subthreshold coefficient $d_{20}^+$ is still absent. Both these effects
are important, as we have shown, and substantially contribute to the
uncertainties.

\vspace{3ex}

In the case of a heavy meson much less is known about the scattering
amplitude of pions off them. At present, the best we could do is to analyze
this at tree level in the heavy meson chiral perturbation theory approach.
At this order, the amplitude can be expressed in terms of the 
pion mass $\mpi$ and the $B^{(*)}$ and $D^{(*)}$ meson masses as well as
the coupling constant $g$ describing the $\pi\HM\HMs$ coupling (with $H=D$ or
$B$). 
The numerical analysis is performed with these as input and shows that:
\begin{itemize}
\item The limited knowledge on the coupling constant $g$ leads to large
  error bars in the numerical results. A higher order calculation of the
  finite volume effect is only worthwhile once this has improved. 
\item Despite our ignorance about the coupling constant $g$ the FVC come
  out to be negligible (well below 1\%) in all regions of the parameter
  space we have explored.
\end{itemize}
Although the practical interest of a calculation of the FVC for heavy
mesons appears to be limited (in view of the fact that they are small), we
found a detailed analysis of some of its features very instructive. We have
discussed, in particular, the difference between a relativistic and a
nonrelativistic version of L\"uscher's formula and analyzed the dependence
of the FVC on the size of the splitting between the pseudoscalar and the
vector heavy meson.

\section*{Acknowledgments}
We are indebted to Bastian Kubis and Christoph Haefeli for useful
discussions at early stages of this work. The Albert Einstein Center for
Fundamental Physics is supported by the ``Innovations- und
Kooperationsprojekt C-13'' of the ``Schweizerische Universit\"atskonferenz
SUK/CRUS''. Partial financial support by the Helmholtz Association through
the virtual institute ``Spin and strong QCD'' (VH-VI-231), by the Swiss
National Science Foundation, and by EU MRTN--CT--2006--035482 (FLAVIA{\it
  net}) is gratefully acknowledged.

\begin{appendix}

\section{Low-energy constants}\label{app:lec}
The LEC $c_i$ and $d_{18}$ are scale independent. For the remaining
LEC the scale independent combinations are given by
\bea
\bar{d}_{16} &=& d_{16}^r(\mu)-\frac{g\left(4-g^2\right)}{8 N
  \Fpi^2}\ln \frac{\Mpi}{\mu}, \nn 
\tilde{e}_1 &=&
e_1^r(\mu)-\frac{\frac{3}{2}g^2-\frac{3}{2}(8c_1-c_2-4c_3)m}{N F^2 m}
\ln \frac{\Mpi}{\mu} \nn
\tilde{e}_3 &=&
e_3^r(\mu)+\frac{1+3g^2+\frac{22}{3}g^4+8c_1m+c_2m-4c_3m}{N F^2
  m}\ln\frac{\Mpi}{\mu},\nn
\tilde{e}_4 &=& e_4^r(\mu)-\frac{10+12g^2+\frac{52}{3}g^4+8c_2m}{N F^2
  m}\ln\frac{\Mpi}{\mu},\nn
\tilde{e}_6 &=& e_6^r(\mu)+\frac{12+8g^2+8g^4}{N
    F^2 m} \ln\frac{\Mpi}{\mu},
\eea 
with $N = 16 \pi^2$.

\section{Details of the fits}\label{app:fit}

The lattice results for the nucleon mass and the axial charge which are used in
the fit are shown in Tables~\ref{tab:mN} and \ref{tab:gA}. We intend not to
strain the chiral expansions too much and do not include lattice results with
pion masses significantly above $350 \MeV$.
\begin{table}[t]
\center
\begin{tabular}{cccc}
$\Mpi [\GeV]$ & $m_N [\GeV] $ & $\delta_{m_N} [\GeV]$ & Collaboration
  \\ \hline
0.13957	&	0.93827 &	0.00008 & physical point  \\  
0.192	&	0.990	&	0.037  & BMW   \\
0.208	&	0.951	&	0.021  & BMW     \\ 
0.271	&	1.067	&	0.030  & BMW     \\
0.307	&	1.132	&	0.031  & BMW   \\ 
0.318	&	1.097	&	0.026  & BMW\\
0.320	&	1.124	&	0.026  & BMW     
\end{tabular}
\caption{Data points for the fit of the nucleon mass, taken from Ref.~\cite{BMW2}.}
\label{tab:mN}
\end{table}
\begin{table}[t]
\center
\begin{tabular}{cccc}
$\Mpi [\GeV]$ & $g_A$ & $\delta_{g_A}$ & Collaboration
  \\ \hline
0.13957	&	1.269	&	0.003 & physical point \\
0.313	&	1.230	&	0.100 & ETM \\
0.350	&	1.210	&	0.070 & RBC/UKQCD  \\
0.352	&	1.250	&	0.060  & LHPC
\end{tabular}
\caption{Data points for the fit of the axial charge. The data points
  are taken from Refs.~\cite{ETM,RBC,LHPC}}.
\label{tab:gA}
\end{table}
 For the remaining quantities, only the value at
the physical point, known from experiment \cite{hohler}, is available,
\bea\label{eq:input}
g_{\pi N} &=& 13.39\pm0.08\co \qquad \Mpi d_{00} = -1.46\pm0.10\co \nn
\Mpi^3 d_{10} &=& 1.14\pm0.02\co \qquad \Mpi^5 d_{20} = 0.200\pm0.005\fs 
\eea
The LEC  $\lbar{i}$ from the mesonic order $p^4$ Lagrangian are not
determined in the fit, but are taken as input parameters without an
error. We use the values given in Ref.~\cite{CGL}. For the pion decay
constant $\Fpi$, the two-loop result from Ref.~\cite{BGT}
with the values of the LEC as discussed in Ref.~\cite{Colangelo:2003hf} is used.
We use the standard $\chi^2$ function 
\bea
\chi^2 &=& \sum_{i,k} \left(\frac{y_k({\bf
    x},M_i)-\bar{y}_{k,i}}{\delta_{k,i}}\right)^2 
\eea
where $y_k({\bf x},M_i)$ is the chiral representation of quantity $y_k$
evaluated at the pion mass $M_i$. The LEC are collected in the vector
${\bf x}$ and $\bar{y}_{k,i}$ denotes the data point of quantity $y_k$
at the pion mass $M_i$ with the error $\delta_{k,i}$. 
The error $\sigma_{y_k}$ of the quantity $y_k$ is given by 
\beq
\sigma_{y_k}^2 = \frac{\partial y_k}{\partial x_l} 
\frac{\partial y_k}{ \partial x_m} C_{lm}
\eeq
with ${\bf C}$ the correlation matrix of the LEC.

\section{Formalism of heavy meson ChPT} \label{sec:HMChPT}

At present there are two frameworks to include heavy mesons into chiral
perturbation theory: a relativistic one introduced in Ref.~\cite{Burdman:1992gh}
and a nonrelativistic one introduced in Ref.~\cite{Wise:1992hn}. We
only present the two Lagrangians very briefly; for a detailed discussion, we
refer to the original publications and to Refs.~\cite{Georgi:1991mr,Manohar:2000dt}.
All conventions we use have been adapted to agree with the latter.
We give the relativistic Lagrangian for fields $B_a^{(Q)}$ and $B_a^{* (Q)}$ that
annihilate a pseudoscalar and a vector meson consisting, respectively, of a
heavy quark $Q$ and a light antiquark $q_a$. From these we build the field

\begin{equation}
	\mathcal{B}_a = \left(\frac{i \slashed{D}_{ab} 
	+ \mH \delta_{ab}}{2\mH}\right) [i B_b \gamma_5 + B_b^{* \mu} \gamma_\mu],
	\qquad \bar{\mathcal{B}}_a = \gamma^0 \mathcal{B}_a^\dagger \gamma^0,
\end{equation}
where we omitted the superscript for simplicity.  The covariant derivative is
given by $D^\mu_{ab} = \partial^\mu \delta_{ab} - \Gamma^\mu_{ba}$,
with the connection
\begin{equation}
	\Gamma_\mu = \inv{2} [ u^\dagger, \partial_\mu u ] = 
						\inv{2} ( u^\dagger \partial_\mu u + u \partial_\mu u^\dagger ) = 
						\frac{i}{4 F^2}\, \varepsilon^{abc} \tau^a \pi^b \partial_\mu
						\pi^c + \mathcal{O}(\pi^4).
\end{equation}
We also need the vielbein
\begin{equation}
	u_\mu = i \{ u^\dagger, \partial_\mu u \} = 
					i ( u^\dagger \partial_\mu u - u \partial_\mu u^\dagger )
					= -\inv{F} \tau^a \partial_\mu \pi^a + \mathcal{O}(\pi^3).
\end{equation}
From these building blocks we can construct the relativistic heavy meson
Lagrangian, which is
\begin{equation} \label{eq:LHMrel}
	\mathcal{L}_{\textit{HM, rel}} = - \mH \: \mbox{Tr} \left[ \bar{\mathcal{B}}_a
	(i \slashed{D}_{ab} - \mH \delta_{ab}) \mathcal{B}_b 
			- \frac{g}{2} \bar{\mathcal{B}}_a \mathcal{B}_b 
			\gamma_\mu \gamma_5 u^\mu \right].
\end{equation}
The second term is not contained in the Lagrangian of Ref.~\cite{Burdman:1992gh} and had
to be added here as it is the source of the one-loop self-energy contributions.

By means of a nonrelativistic reduction procedure similar to the one used
in heavy-baryon chiral perturbation theory \cite{Jenkins:1990jv,
Bernard:1992qa}, we can derive the Lagrangian of Ref.~\cite{Wise:1992hn} from this. Instead of
$\mathcal{B}_a$, we use the field
\begin{equation}
	H_a = \frac{\slashed{v} + 1}{2} [i P_a \gamma_5 + P_a^{* \mu}\gamma_\mu],
\quad
	\bar{H}_a = \gamma^0 H_a^\dagger \gamma^0 = [i P_a^\dagger \gamma_5 + P_a^{*
\mu \dagger} \gamma_\mu] \frac{\slashed{v} + 1}{2},
\end{equation}
where $v^\mu$ is the four velocity of the heavy meson and the fields $P_a$ and
$P_a^{* \mu}$ replace $B_a$ and $B_a^{* \mu}$ respectively. Note that the new fields
differ in normalization by a factor $1/\sqrt{\mH}$, which ensures that the
nonrelativistic Lagrangian does not depend on \mH. Inserting these fields
into Eq.~(\ref{eq:LHMrel}) and expanding to leading order in $1/\mH$, we find
the nonrelativistic Lagrangian
\begin{multline}
	\mathcal{L}_{\textit{HM, nrel}} = -i \: \mbox{Tr} 
				\left( \bar{H}_a v_\mu \partial^\mu H_a \right) \\
				+i \: \mbox{Tr} \left( \bar{H}_a\, H_b v_\mu \Gamma^\mu_{ba} \right)
				+ \frac{g}{2} \: \mbox{Tr} \left( \bar{H}_a\, H_b \gamma_\mu \gamma_5
u_{ba}^\mu \right).
\end{multline}
Up to the phase conventions, this agrees with the nonrelativistic Lagrangian of
Ref.~\cite{Wise:1992hn}.

\end{appendix}

 \end{document}